\documentclass[sigconf]{acmart}

\usepackage{lineno,hyperref}
\modulolinenumbers[20]
\usepackage[titlenumbered,ruled]{algorithm2e}
\usepackage{graphicx, subfigure}
\usepackage{color}
\usepackage{booktabs}
\usepackage{longtable}
\usepackage{lipsum}
\usepackage{supertabular}

\begin{document}

\copyrightyear{2018} 
\acmYear{2018} 

\title{Automatic Investigation Framework for Android Malware Cyber-Infrastructures}

\author{ElMouatez Billah Karbab}
\affiliation{%
  \institution{Concordia University}
}

\email{e\_karbab@encs.concordia.ca}
\author{Mourad Debbabi}
\affiliation{%
  \institution{Concordia University}
}
\email{debbabi@encs.concordia.ca}

\begin{abstract}
The popularity of Android system, not only in the handset devices but also in IoT devices, makes it a very attractive destination for malware. Indeed, malware is expanding at a similar rate targeting such devices that rely, in most cases, on Internet to work properly. The state-of-the-art malware mitigation solutions mainly focus on the detection of the actual malicious Android apps using dynamic and static analyses features to distinguish malicious apps from benign ones. However, there is a small coverage for the Internet/network dimension of the Android malicious apps. In this paper, we present \textsf{ToGather}, an automatic investigation framework that takes the Android malware samples, as input, and produces a situation awareness about the malicious cyber infrastructure of these samples families. \textsf{ToGather} leverages the state-of-the-art graph theory techniques to generate an actionable and granular intelligence to mitigate the threat imposed by the malicious Internet activity of the Android malware apps. We experiment \textsf{ToGather} on real malware samples from various Android families, and the obtained results are interesting and very promising.
\end{abstract}

\maketitle

\section{Introduction}

Mobile devices are important gadgets in our lives. Nowadays, mobile systems, especially Android, and their applications (apps) dominate most of our daily economic and social interactions. Android has the biggest share in the mobile computing industry \cite{idc2016smartphone} due to its open-source distribution and sophistication. Besides, it became not only the dominant platform for mobile phones and tablets, but is also gaining increasing attention and penetration in the realm of Internet of Things (IoT) \cite{android_wear, brillokey}. The ubiquitous nature and popularity of Android OS made it the first target of malicious threats in mobile computing platforms \cite{GData_2015_stats}. Indeed, malware apps are not only targeting conventional devices such as phones and tablets, but also more critical systems such as home IoT devices. The latter could allow the adversary to achieve more severe attacks, which could inflict physical damages \cite{android_auto} as the attacker could gain access to physical system controllers of cars, air conditioning systems, refrigerators, etc. 

Mobile and IoT devices are more critical than personal computers in many ways: (i) In contrast with personal computers, they are equipped with sophisticated sensors, from cameras and microphone to gyroscope and GPS. These various sensors open a whole new world of applications for end-users. However, they also unleash unprecedented potential cyber-threats that could be committed by adversaries who gain access to these resources through Android malware apps. (ii) Thin devices (smart handsets and IoT devices) have limited resources in terms of computation, energy power, and network bandwidth compared to PCs. This makes extensive security analyses very expensive, if not impossible in some cases, to track maliciousness indicators whether dynamic or static in nature. Therefore, the adversary needs less sophisticated malicious apps compared to PC ones to achieve the attack. (iii) A thin device could inflict more damage than a PC due to its high portability, and hence could infect/damage a large number of networks (e.g., work, home, restaurant, airport). Indeed, infected thin  devices could play the role of a payload transporter to harm other systems and networks. (iv) In terms of deployment, the number of thin devices (including Android ones) is by far larger than the number of PCs. Therefore, an adversary that leverages malicious apps could infect more IoT devices than PCs. The attacker could infect and control (tens of) thousands of PCs and use them as her/his malicious cyber-infrastructure. Nowadays, malicious cyber-infrastructures could reach (tens of) millions of devices if we include devices in TVs, smart watches, connected cars, etc. (v) Finally, the centralized mechanism through which Android apps are distributed using App repositories \cite{google_play} allows for the distribution of malicious apps that bypass vetting systems and hence be available on a huge number of end-user devices.

The aforementioned factors highlight the urgent need to design and implement new methodologies, techniques and  tools to mitigate cyber-threats against Android mobile and IoT devices, especially that we are witnessing a convergence between Android and IoT devices. IoT devices could run Android OS or a lightweight version of it. In this context, Google proposes AndroidThing  \cite{brillokey}, an Android-based IoT operating system. On the other hand, Android may animate other IoT devices that control systems such as smart homes or smart buildings. To mitigate these cyber-threats  \cite{GData_2015_stats}, we need to have an accurate situational awareness of the threat landscape. The state-of-the-art Android security solutions mainly concentrate on: (i) Static analysis \cite{arp2014drebin, karbab2016dna, feng2014apposcopy, yang2014apklancet}, where the emphasis is on the actual Android malware file (Android Packaging APK): Here, the community tends to fingerprint Android malware using approximate fingerprints or learning models that leverage statistical features engineered from the static content. Static analysis is not generally effective in the presence of obfuscation techniques. (ii) Dynamic analysis \cite{canfora2016acquiring, spreitzenbarth2013mobile, ali2016aspectdroid, zhang2013vetting} based on the reports generated after executing the actual malware in a sandboxing system: The security analysis leverages these reports to discover and fingerprint malicious behaviours of Android malware samples. The dynamic analysis tends to be more resilient against obfuscation. However, it is more time consuming compared to static analysis. (iii) Hybrid approaches \cite{yuan2014droid, grace2012riskranker,bhandari2015draco, vidas2014a5} leverage both static and dynamic analyses techniques to achieve a higher detection performance.

However, current Android malware solutions do not address the network dimension of mobile and IoT security. Furthermore, a common important characteristic between IoT devices and smart handsets is Internet access. Therefore, having malicious apps could allow the adversary to connect to infected devices at any time. Besides, Internet access is far from being a suspicious permission in Android vetting system. More precisely, the gap resides in the lack of situational awareness about malicious cyber-infrastructures that relate Android malware apps and their families. By cyber-infrastructure, we mean all the domains and IP addresses, i.e., the network information that is used by the adversary to control, download, upload, or at least, collect sensitive information through malicious apps that are already installed on infected Android devices (e.g. smart handset and IoT devices). Solutions, such as \cite{boukhtouta2015graph} \cite{nadji2013connected}, address malicious infrastructures in general focusing on malware samples and their families but without a special emphasis on Android-based platforms.  In other words, there is a need for online solutions that leverage the large number of detected Android malware samples from different families. The latter should be the starting point of security solutions to achieve a situational awareness about malicious cyber-infrastructures underlying daily Android malware at different granularity levels. In other terms, the intention is to achieve a better understanding and focus on the malicious cyber-infrastructures underlying one Android malware sample, one malware family of samples, or several families at the same time.

In this respect, we propose \textsf{ToGather}, an automatic investigation framework for Android malware cyber infrastructures. \textsf{ToGather} framework is a set of techniques and tools together with security feeds to automatically achieve a situational awareness about Android malware. Actually, \textsf{ToGather} characterizes the cyber-infrastructure of a given malware sample, a set of samples, family or families as a multipartite graph that relates malware samples and the corresponding network footprint in terms of IPs and domains. \textsf{ToGather} goes even a step further by dividing this cyber-infrastructure into sub-infrastructure components based on the connectivity between the nodes. The result is in fact multiple network communities representing many sub-cyber-infrastructures that are related to the Android malware sample or family. To this end, \textsf{ToGather} leverages the enormous amount of cyber-threat intelligence  that is derived from various sources such as spam, Windows malware, darknet, and passive DNS to ascribe cyber-threats to the corresponding cyber-infrastructure. Accordingly, the input of \textsf{ToGather} framework is made of malware samples, and the output is networks of cyber-infrastructures together with their network footprint, which would give the security practitioner an overview of Android malware cyber-activities on the Internet.

The process of \textsf{ToGather} framework starts by taking, as input, Android malware samples. First, we extract network information from these malicious apps. For this purpose, we use a hybrid approach,  where both static and dynamic analyses are applied on malware. The resulting network information (IPs and domain names) of the malware sample represents the malicious nodes of its malicious cyber-infrastructure. However, the network information could be very noisy, as it might include several benign domain names of well-known sites, and the same applies to IP addresses. Hence, \textsf{ToGather} filters these entries through whitelisting in order to remove such IPs and domain names. Afterwards, \textsf{ToGather} correlates the network footprint with a passive DNS database to enrich the network information in two ways: (i) Get the IP addresses resolved from the current domain names list. (ii) Get the domain names that point to the collected IP addresses. The result is an enriched network information that has more coverage in terms of malicious cyber activity of the input malware samples.  However, this information could be richer if we structure it; hence, \textsf{ToGather} builds from the network information a multi-partite graph connecting the hashes of malware samples to the corresponding IP addresses and domain names. The heterogeneous graph is used to derive abstract homogenous graphs where the emphasis is put on the network information while abstracting away from the malware hashes (since they have the same family in a typical use case).  The homogeneous graphs, namely threat networks, represent cyber-infrastructures of  Android malware. \textsf{ToGather} applies a highly scalable community detection algorithm \cite{fast08blondel} on this threat network to extract sub-threat networks with high connectivity aiming to give a more granular view to the security practitioner. Besides, we apply page ranking algorithm on these sub-cyber-infrastructures in order to rank the nodes (information network) according to their importance in terms of connectivity among sub-graphs. This indeed results in  actionable intelligence that could be leveraged for instance to take-down operations. Finally, for each sub-threat network, we correlate the resulting cyber-infrastructure with well-known malicious information networks to label the underlying malicious activities. \textsf{ToGather} framework automates the previous steps to help security analysts achieve a great deal of situational awareness on Android malware and its activities on the Internet. As such, our contribution is essentially the framework as a whole and not only the components. 

The main contributions of this paper are: 

\paragraph{(1)} We design and implement \textsf{ToGather}, a simple, yet practical framework to generate a granular situational awareness report on the malicious cyber infrastructures underlying Android malware.

\paragraph{(2)} We propose a correlation mechanism with multiple cyber-threat intelligence feeds, which enrich, not only the resulting malicious cyber-infrastructure intelligence, but also the labeling of the tracked malicious activities.

\paragraph{(3)} We evaluate \textsf{ToGather} framework on real Android malware samples from Drebin malware dataset. The evaluation shows promising and interesting findings.

\section{Overview}

\subsection{Threat Model}

We position \textsf{ToGather} as a detector of malicious cyber-infrastructures of Android malware. It is designed to uncover threat networks and the sub-networks from a seed of Android malware samples. However, malware detection is described in existing proposals \cite{arp2014drebin, karbab2016dna, canfora2016acquiring, spreitzenbarth2013mobile,yuan2014droid, grace2012riskranker} and is out of the scope of this paper. \textsf{ToGather} does not guarantee zero false-positives due to the large number of benign domain names and IP addresses that might not be filtered out with \textsf{ToGather} whitelists. \textsf{ToGather} is very resilient to obfuscation during the extraction of the network information from Android malware because it applies both static and dynamic analyses. Hence, if the static content is heavily obfuscated, \textsf{ToGather} is still able to collect IP addresses and domain names from dynamic analysis reports. 

\subsection{Usage Scenarios}

\textsf{ToGather} is designed to be practical and efficient in the hands of security practitioners.

\begin{itemize}

\item Security analyst uses \textsf{ToGather} framework as an investigation tool to minimize the efforts of generating threat networks for a given Android malware family. The analyst leverages the IP addresses and domain names ordered by their importance in the generated threat network to prioritize the takedown and mitigation operations.

\item \textsf{ToGather} acts as a monitoring system. It analyzes a malware feed of Android malware family (e.g., new samples on a daily basis) to generate a snapshot for threat network and uncover the malicious activities (spam, phishing, scanning, and others). Periodic reporting gives insights into the cyber evolution and the malicious behaviors of a given malware family over time.

\item \textsf{ToGather} measures the Android malware activity on top of cloud vendors by reporting that a given Android malware family is using a specific cloud vendor infrastructure for its malicious activity during a period of time.

\end{itemize}

\section{Methodology}
In this section, we present the overall workflow of \textsf{ToGather} framework, as shown in Figure \ref{fig:overview}, starting from the Android malware samples ending with the produced relevant security intelligence:

\begin{figure*}[!htb]
  \centering
      \includegraphics[width=0.99\textwidth]{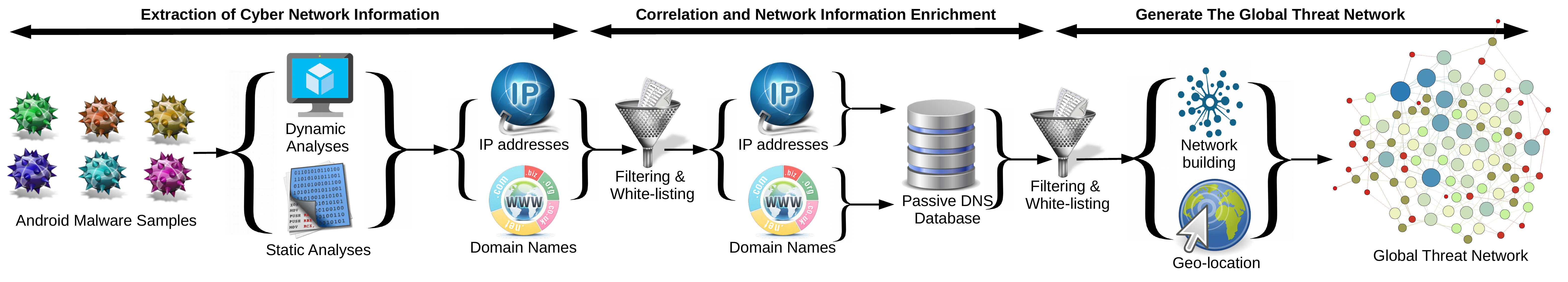}
  \caption{Approach Overview}
  \label{fig:overview}
\end{figure*}


1) The first step in \textsf{ToGather} framework consists of deriving network information from Android samples in a given window analysis (e.g., day, week, month) whether they are from the same malware family or not. However, we consider one malware family as a typical use-case of \textsf{ToGather}, as presented in the evaluation section. \textsf{ToGather} conducts dynamic and static analyses where each analysis produces a report for each Android malware sample. Therefore, a given malware has two reports from dynamic and static analyses. Leveraging both analysis types enhances the resiliency of \textsf{ToGather} against common obfuscation techniques. The latter aims to hide relevant information about malicious activities such as domain names and IP addresses (network information). Afterwards, \textsf{ToGather} extracts network information strings from the analysis reports. At this point, we apply a simple text pattern search to find IP addresses and domain names. In static analysis, we mainly concentrate on the Dalvik compiled code (classes.dex) for the extraction. We collect network information more efficiently from dynamic analysis reports since they are more structured and have labeled fields.

2) Next, we filter the extracted network identifiers from noise information such as non-routed IP addresses. Also, we filter domain names and URLs that use Unicode characters. For the current \textsf{ToGather} implementation, we do not consider domain names and URLs written in other languages such as Japanese or Arabic. In the case of URL links, we keep only domains. To this end, we have a set of valid IP addresses and domain names found in Android malware. It is important to notice here that each network information is tagged by the underlying malware hash and this tag will be kept during all the workflow steps of \textsf{ToGather}. To minimize false positives, \textsf{ToGather} applies whitelisting mechanisms. For domain names, \textsf{ToGather} leverages the complete Alexa \cite{alexatop_web} and Quantcast \cite{quantcast_web} (more than one million domain name). However, the number of white domain names is a hyper-parameter in \textsf{ToGather} that we could use to control the amount of false positives. In the case of IP addresses, we leverage a set of public white IPs such as Google DNS servers and other ones \cite{tracemyip_web}. It is important to stress that \textsf{ToGather} considers public cloud vendor IPs and domain names as a whitelist. The aim is to observe and then gain insight into the use the cloud infrastructure by Android malware. This idea originates from the observation that Android malicious apps (and malware in general) make more use of the cloud as a low-cost infrastructure for their malicious activity.

3) In this step, we propose a mechanism to enhance and enrich the malicious network information to cover related domains and IPs. In essence, \textsf{ToGather} aims at answering the following questions: (i) What are the IP addresses of current malicious domains? Here we investigate the IP addresses of server machines that host malicious activities. The latter are most likely related to the analyzed Android malware. (ii) What are the domain names pointing to the current malicious IP addresses? The intuition is that a malicious server machine with a given IP address could host various malicious contents and the adversary could use multiple domains pointing to such contents. To answer this question, \textsf{ToGather} has a module to enrich network information by using passive DNS replication.  The latter is a technology that builds zones replicate without the cooperation from zone administrators, based on captured name server responses, as presented in Section \ref{sec:pdns}. We use the network information, whether IP addresses or domains, as parameters to two functions applied on a passive DNS database. The goal of the function is to enrich the list of domains and IP addresses that could be part of the adversary threat network. The enrichment services are: (i) GetIP(Domain): This function takes a domain as a parameter to query passive DNS database. The result is all IP addresses the domain points to. (ii) GetDomain(IP): This function gets all the domains that resolves to the IP address given as a parameter.

We consider passive DNS correlation for two reasons: (i) A small number of Android malware samples generally yields limited network information.  (ii) Security practitioners aim at having a more comprehensive situational awareness about malware Internet activity. As such, they would like to consider all related IPs and domain names. The result of the correlation is a set of IP addresses and domains enriched using passive DNS, related to Android malware apps. The correlation results could, however, overwhelm the investigation process. Passive DNS correlation is therefore optional if we have a big number of samples from a given Android family. \textsf{ToGather} applies network information enrichment using the passive DNS replication. The correlation with passive DNS could produce some known benign entries. For this reason, we filter the likely harmless network information by matching the newly found ones against top Alexa \cite{alexatop_web} and Quantcast \cite{quantcast_web} domain names and known public IP addresses \cite{amazonip_web}.

5) At this stage, we have a set of network information tagged by malware hashes. To extract relevant and actionable intelligence, \textsf{ToGather} aggregates all the previous records into a heterogeneous network with different types of nodes: \textit{malware hashes}, \textit{IP addresses} and \textit{domain names}. We consider the heterogeneous network that is extracted from a given Android malware family as the malicious activity map of that family on the Internet. We call such a heterogenous network, a \textit{threat network}. Furthermore, \textsf{ToGather} produces homogenous networks by executing multiple projections according to the node type (IP address or domain name). For example, in the IP projection, the projection of a malware hash connecting two IP addresses would be only the two IPs connecting to each other. Therefore, \textsf{ToGather} produces three homogeneous graphs, one only considers IP addresses connections, and the other only considers domain names connections. Finally, we consider a threat network with  IPs and domains as one type network information. The goal of homogenizing the connection network is to apply graph analyses that need the graph homogeneity (i.e., IP threat network), domain threat network, and network information threat network.

\begin{figure*}[!htb]
  \centering
      \includegraphics[width=0.95\textwidth]{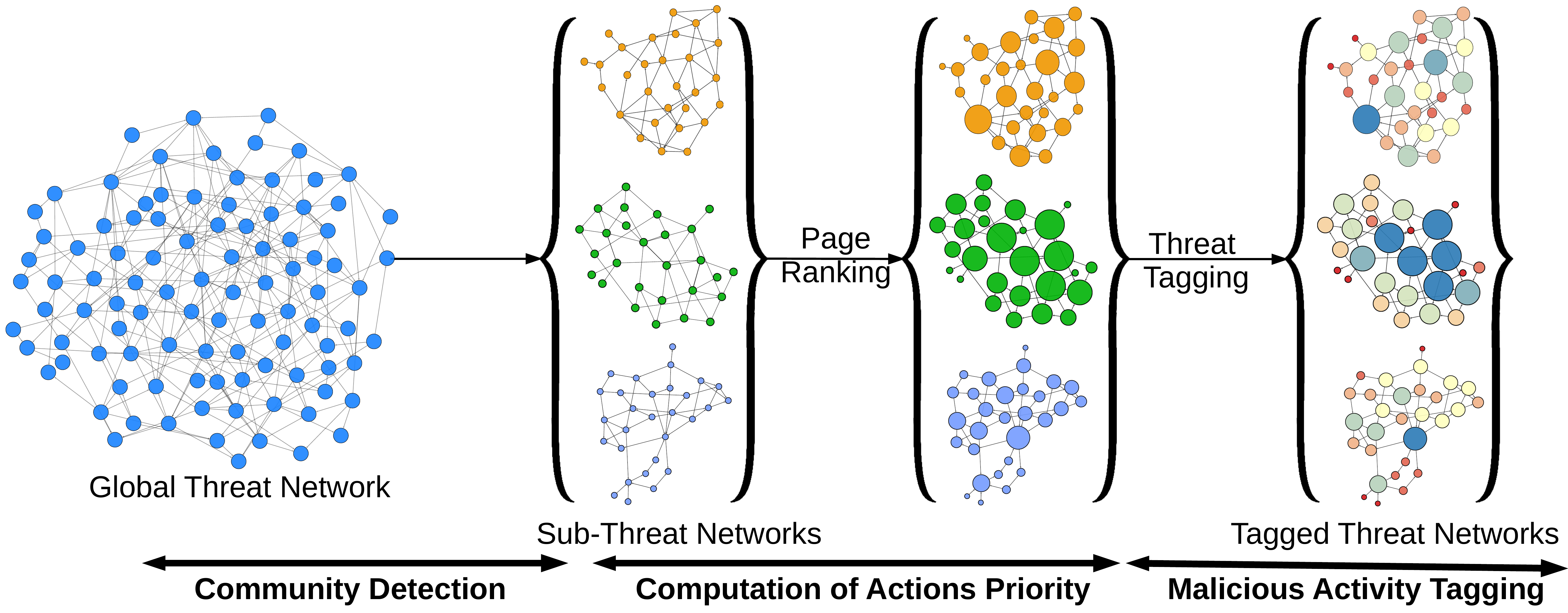}
  \caption{Graph Analysis Overview}
  \label{fig:graph_overview}
\end{figure*}

6) Further, \textsf{ToGather} aims at producing more granular graphs (see Figure \ref{fig:graph_overview}) from the generated threat networks derived in the previous step. In this respect, \textsf{ToGather} checks the possibility of community identification in these threat networks based on the connectivity between nodes. The higher is the connectivity between the nodes in a particular area of the network, the more is the possibility to have a malicious community in such area. For community detection (Section \ref{sec:community_detection}), we adopt a highly scalable algorithm \cite{fast08blondel} to enhance \textsf{ToGather} community detection module. The intuition behind using the community concept is: (i) Considering \textsf{ToGather} typical usage scenario, where we enter Android malicious apps from the same family, the community could define different threat networks that are related to the malicious activities. In other words, either one adversary is using these threat networks as backups or we have instead multiple adversaries. In the case of Android malware, the second hypothesis is more plausible because of the cheap repackaging of existing malware samples to suit the need of the perpetrator. (ii) In case \textsf{ToGather} receives Android malware from different families, the communities could be interpreted as the threat networks of different Android malware families to focus on the relation between them. The output of this step is a set of threat networks related to IPs, domains, and network information and their communities (sub-threat networks).

7) To produce actionable cyber-threat intelligence, we leverage Google page ranking algorithm (Section \ref{sec:pageranking}) to produce ranking scores for critical nodes of a given (sub) threat network. Consequently, the investigator would have some priority list when it comes to mitigation or take down of nodes that are associated with a malicious cyber-infrastructure. As a result, \textsf{ToGather} produces each (sub) threat network of the Android malware family together with the ranking of each node. Because \textsf{ToGather} generates multiple homogeneous graphs based on the node type (IP, domain, network information), it produces different ranking lists based on the node type. Therefore, the security practitioner will have the opportunity of selecting the node type when executing the mitigation or the take down to protect his system. In such case, an IP node could be more suitable as it could be blacklisted for instance. Also, it is important to mention that it is expensive for the adversary to get new IP addresses. In contrast, domain names could be frequently changed due to their affordability.

8) We do not focus only on Android malware. Instead, we aim at gaining insights into the shared network IP and domains with other platform malware families. Indeed, the adversary tends to have many malicious weapons in several operating systems to achieve the maximum coverage. Therefore, similarly to the first step, we conduct dynamic and static analyses on Windows and Linux malware samples to extract the corresponding network information. The same step is applied to this network information. Afterwards, we correlate the Android network information with the non-Android malware information to discover another dimension of the adversary network. The result will be all the IP addresses and the domains of Android malware in addition to all network records of a given non-Android malware family if they share some network information. It is important to notice that the information networks of non-Android malware are also labeled by malware families. Therefore, the result of this step is the previous (sub) threat networks tagged by Android malware family in addition to tags of other platform malware. So, the security analyst would have a clearer view on the Android cross-platform malicious activity.

9) In this final step of \textsf{ToGather} workflow, we leverage another cyber-threat source, namely sub-threat networks, to label malicious activities that are committed by the produced communities. Specifically, we leverage network information that is collected from different security data. The current \textsf{ToGather} implementation includes the correlation with spam emails, reconnaissance traces and phishing URLs. Therefore, the investigator will  not only have the cyber-infrastructure of the Android malicious family but also if it is part of other cyber malicious activities that are conducted by the infrastructure.  

We consider \textsf{ToGather} as an active service that receives at every epoch time (day, week, month) Android malware with its corresponding family (the typical use case) and produces valuable intelligence about this malware family.


\section{Threat Communities Detection} \label{sec:community_detection}
\begin{figure}[!htb]
  \centering
      \includegraphics[width=0.48\textwidth]{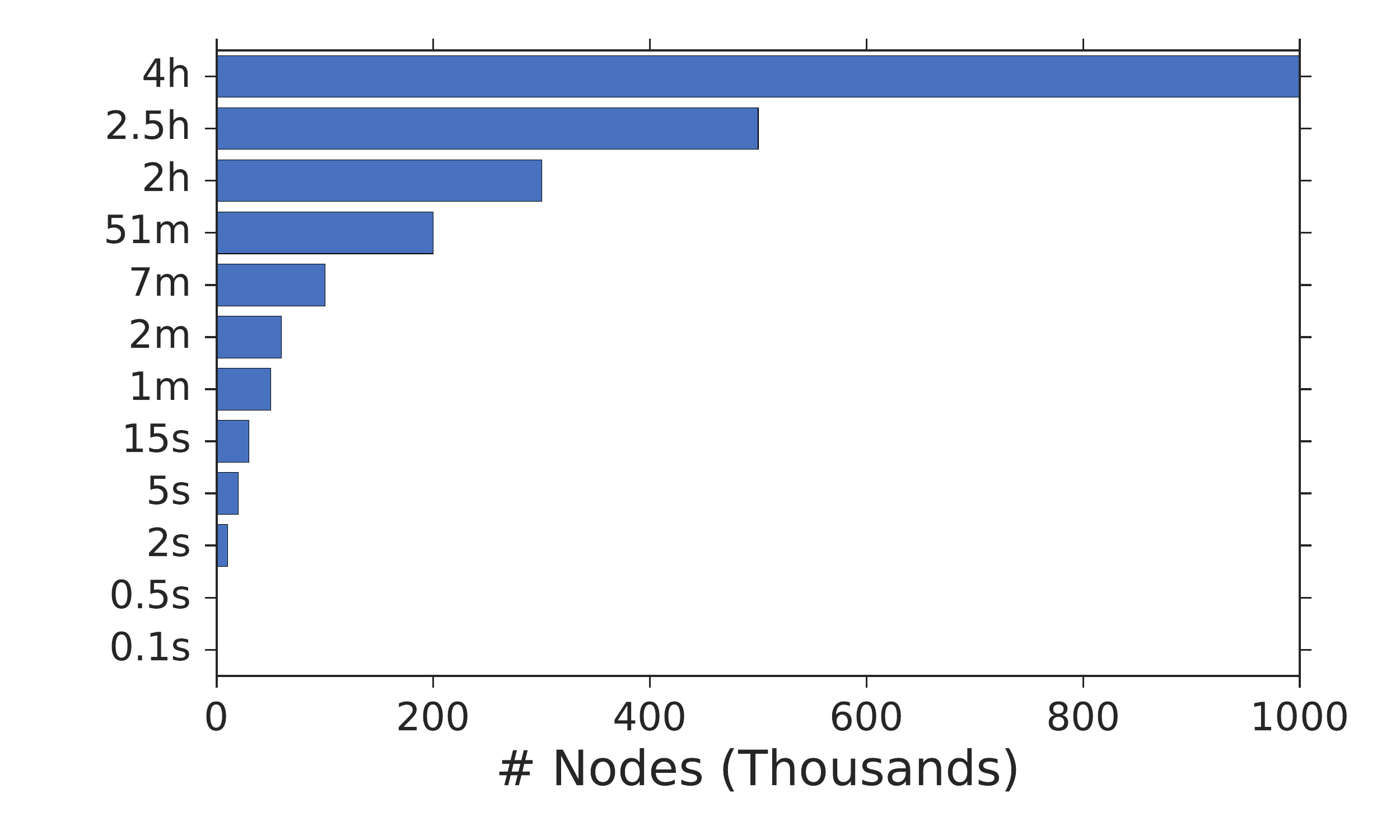}
  \caption{Scalability of the Community Detection}
  \label{fig:scale_million}
\end{figure}

A scalable community detection algorithm is essential to extract communities from the threat network. For this reason, we empower \textsf{ToGather} with the Fast Unfolding Community Detection algorithm \cite{fast08blondel}, which can scale to billions of network links. The algorithm achieves excellent results by measuring the \textit{modularity} of communities. The latter is a scalar value $M \in [-1, 1]$ that measures the density of edges inside a given community compared to the edges between communities. The algorithm uses an approximation of the modularity since finding the exact value is computationally hard \cite{fast08blondel}. Our main reason to choose the algorithm proposed in \cite{fast08blondel} is its scalability. As depicted in Figure \ref{fig:scale_million}, we apply the community detection on a million-node graph with a medium density ($P=0.001$ probability of node A is another node B), which we believe has a similar density to the threat network generated from Android malware samples. For the sake of completeness, we perform the same experiment on graphs with a different probability $P$. As presented in Figure \ref{fig:scale_vhigh},  we are able to detect communities in $30,000$-node graphs with ultra density (unrealistic) in a relatively small (compared to the time dedicated to the investigation) amount of time ($3 hours$).

\begin{scriptsize}
\begin{figure}[ht!]
     \begin{center}
        \subfigure[$P=0.001$ Medium]{%
            \label{fig:scale_meduim}
            \includegraphics[width=0.22\textwidth]{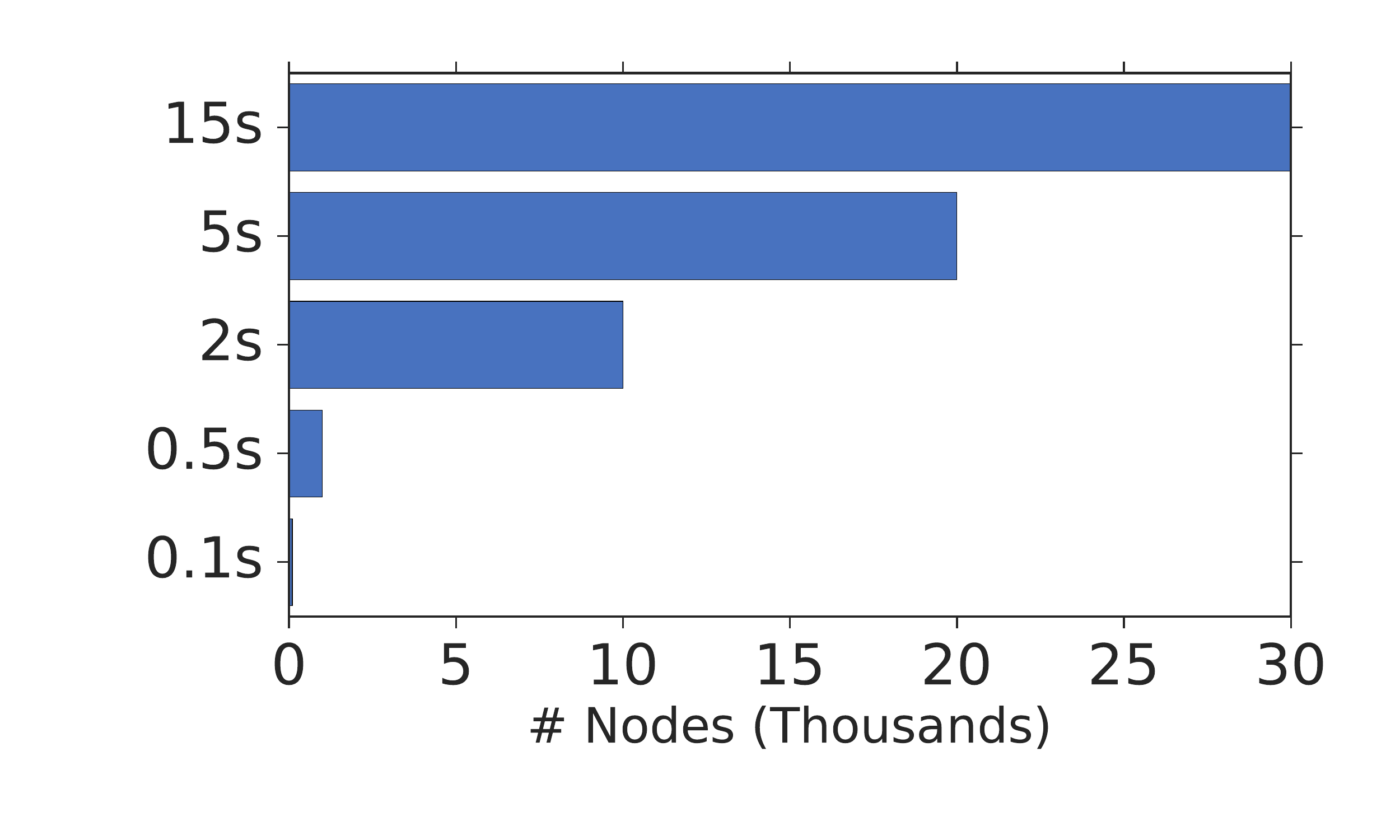}
        }
        \subfigure[$P=0.01$ High]{%
           \label{fig:scale_high}
           \includegraphics[width=0.22\textwidth]{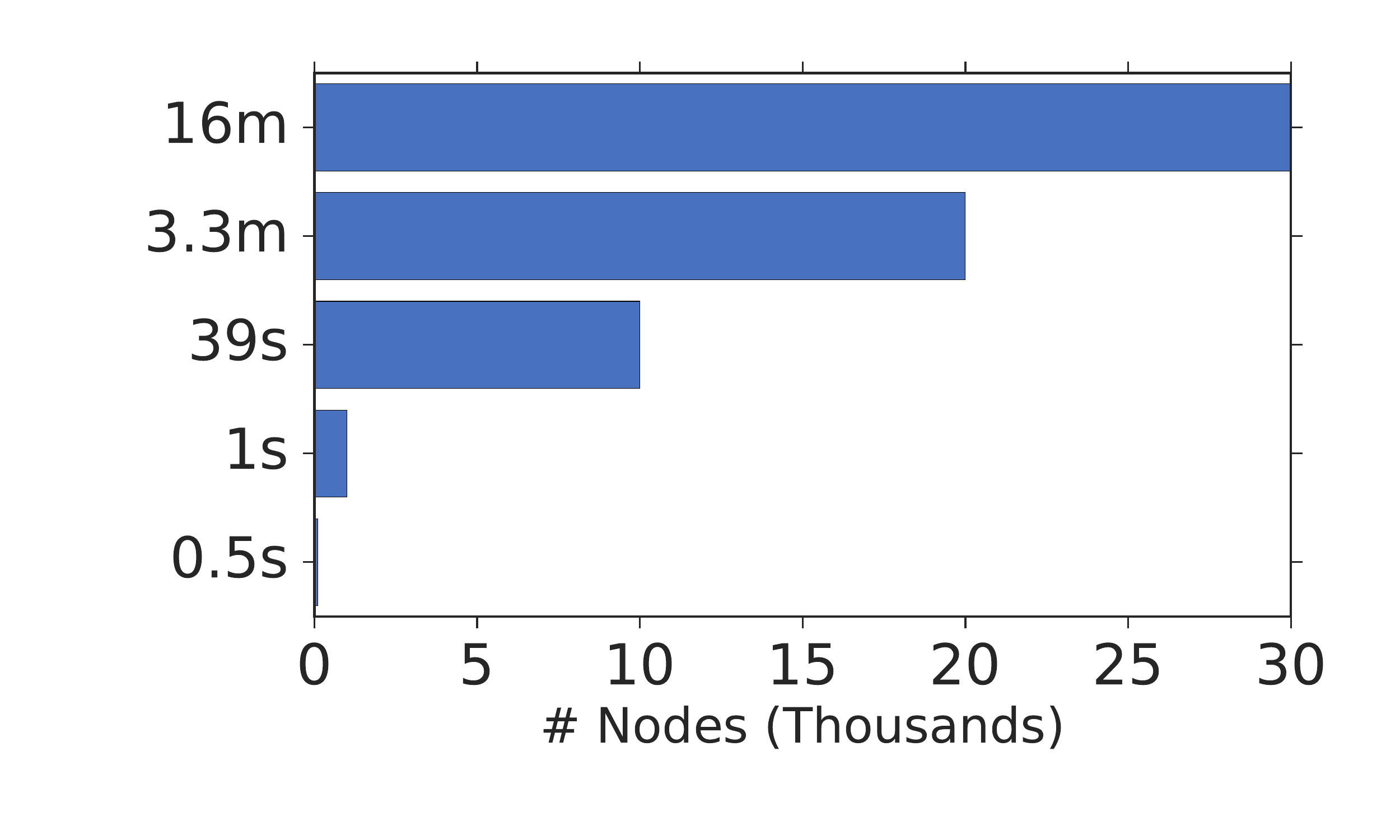}
        }
        \subfigure[$P=0.05$ Very High]{%
            \label{fig:scale_vhigh}
            \includegraphics[width=0.22\textwidth]{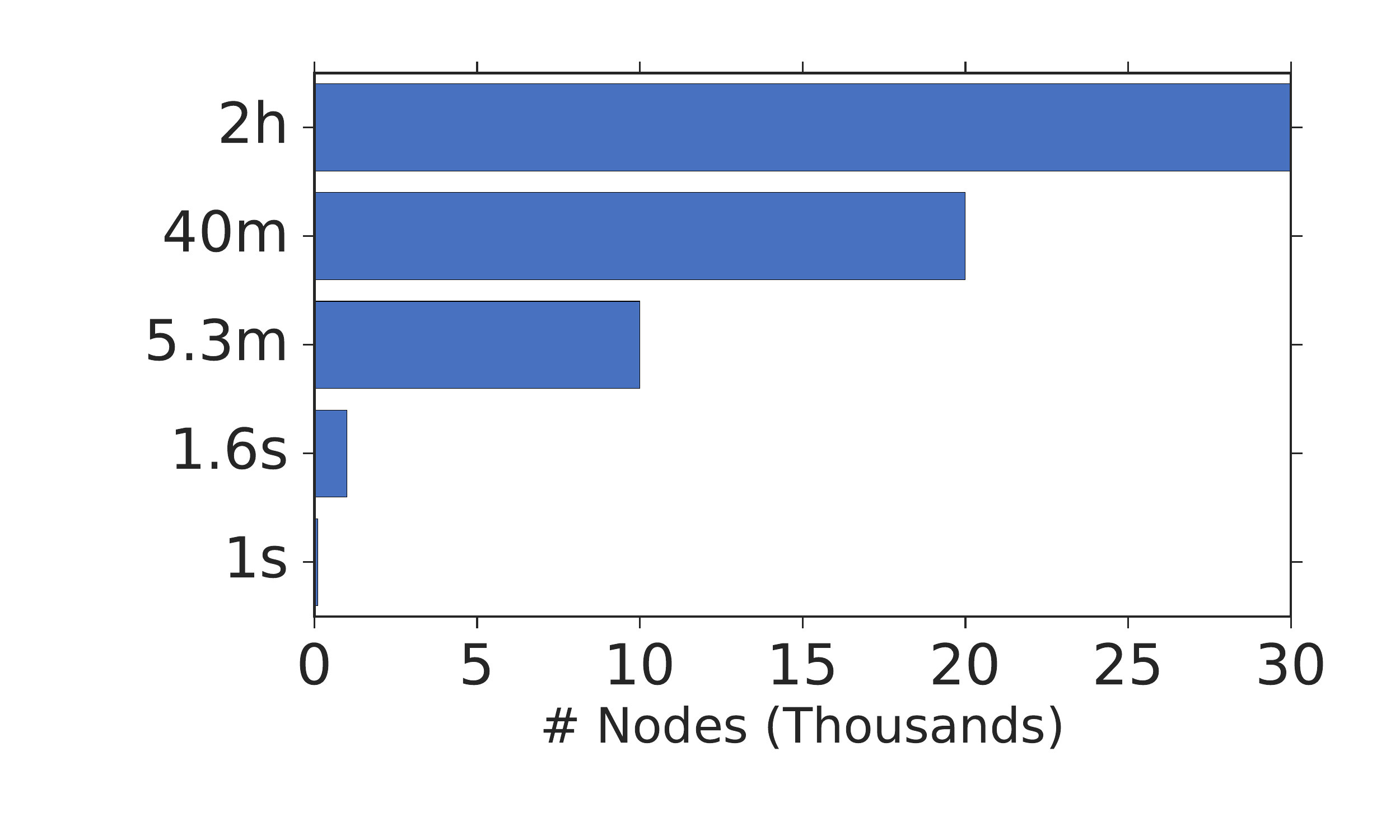}
        }
        \subfigure[$P=0.10$ Ultra High]{%
           \label{fig:scale_ultra}
           \includegraphics[width=0.22\textwidth]{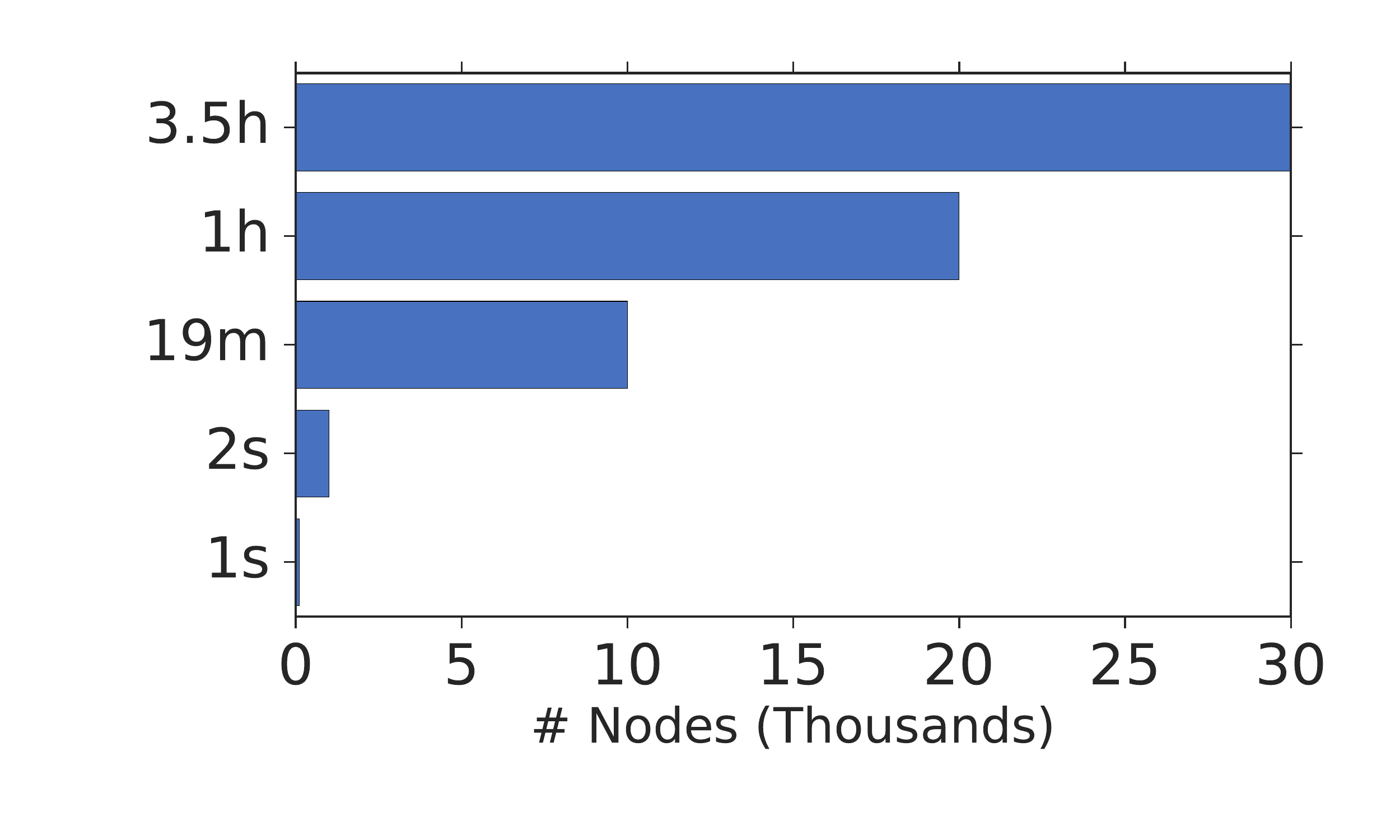}
        }
    \end{center}
    \caption{
        Graph Density Versus Scalability 
     }
   \label{fig:scale_vs_desity}
\end{figure}
\end{scriptsize}

The previous algorithm requires a homogeneous network, as input, to work correctly. In our case, the threat network generated from the network information is a heterogeneous network because it contains two main node types: (i) The malware sample identifier, which is the cryptographic hash of the malware sample. (ii) The network information: the domain names and the IPv4 addresses. In the current implementation, we do not consider IPv6 addresses and domain names in other languages. Also, we apply the projection on the first heterogeneous network to generate homogeneous graphs. To do so, \textsf{ToGather} makes the graph projection by abstracting away from the malware identifier and only takes the network information, i.e., if the malware identifier connects to two IPs, the projection would produce only the two IPs connecting to each other. To this end, we get different projection results based on the node abstraction: (i) General threat network contains both IP addresses and domain names. (ii) IP threat network contains only IP addresses. (iii) Domains threat network contains only domain names. 

Furthermore, \textsf{ToGather} could mine sub-threat networks that have highly connected nodes compared to the rest of the cyber-threat network. The intuition here is that each sub-threat network could be a different malicious infrastructure that is used by an adversary. The security practitioner could automatically segregate possible cyber-infrastructures that could lead to different attacks even if we use only one Android malware family. To achieve such scenario, we apply the previous community detection algorithm on the different threat network to check for possible sub-graphs. Also, \textsf{ToGather} filters nodes (IPs, domains) with weak links to others nodes, as we interpret them as false positives (leaves or parts of tiny sub-graphs).

\section{Actions Prioritization}\label{sec:pageranking}
From the community detection,  \textsf{ToGather} checks if there is possible sub-graphs in the threat networks based on node connectivity. Even though the sub threat networks zoom into malicious cyber-infrastructures of a given Android malware family, the security practitioner could not mitigate against the whole threat network at once. For this reason, \textsf{ToGather} proposes an action priority system. The latter takes the IP, domain or both, and threat network  and produces an action priority list based the maliciousness of each node. By leveraging the graph structure of the threat network, we measure the maliciousness of a given node by its degree, meaning, the number of edges that relate it to other nodes.  From a security point of view, the more connections an IP or domain has, the more it is important for a malicious cyber-infrastructure. Therefore, our goal is to build a priority list sorted by the damage, an IP or a domain, which can inflict  in terms of malicious activity. The importance of nodes in a network graph is known as \emph{node's centrality}. The latter represents a real-valued function produced to provide a ranking, which identifies the most relevant nodes (\cite{borgatti2005centrality}). For this purpose, some algorithms have been defined, such as Hypertext Induced Topic Search (HITS) algorithm (\cite{kleinberg1999authoritative}) and Google's PageRank algorithm (\cite{brin1998anatomy}). In our approach, we adopt Google's PageRank algorithm due to its efficiency, feasibility, less query time cost, and less susceptibility to localized links (\cite{nidhi2012comparative}). In the following, we briefly introduce the PageRank algorithm and the random surfer model.

\subsection{PageRank Algorithm}

\begin{definition}(PageRank).
Let $I(v_i)$ be the set of vertices that link to a vertex $v_i$ and let $deg_{out}(v_i)$ be the out-degree centrality of a vertex $v_i$.  The PageRank of a vertex $v_i$, denoted by $PR(v_i)$, is provided in Eq. $1$:

\begin{equation}
PR(v_i) = d \left[\sum_{v_j\in I(v_i)}{\frac{PR(v_j)}{deg_{out}(v_i)}}\right] + (1-d)\frac{1}{|D|}
\end{equation}

\end{definition}

The constant $d$ is called \emph{damping factor}, assumed to be set to $0.85$ \cite{brin1998anatomy}. Eq. $1$ produces one equation per node $v_i$ with an equal number of unknown $PR(v_i)$ values. The PageRank algorithm tries to find out iteratively different PageRank values, which sum up to 1 ($sum_{i=1}^{n} PR(v_i)=1$). The authors of the PageRank algorithm considers the use case of web surfing,  where the user starts from a web page and randomly moves to another one through a web link. If the web surfer is on page $v_j$ with a probability or a damping factor $d$, then the probability to change page $v_i$ is  $\frac{1}{deg_{out}(v_j)}$. The user could follow the links and teleport to a random web page in $V$ with $1-d$ probability. The described surfing model is a stochastic process, and $W$ is a stochastic transition matrix, where node ranking values are computed as presented in Eq. $2$:

\begin{equation}
\vec{PR} = d \left[W . \vec{PR}\right] + (1-d)\frac{1}{|D|}\vec{1}
\end{equation}

\noindent The stochastic matrix $W$ is defined as follows:\\

\indent $w_{ij}=\frac{1}{deg_{out}(v_j)}$ if a vertex $v_j$ is linked to $v_i$ \\
\indent $w_{ij}=0$ otherwise \\

The notation $\vec{R}$ stands for a vector where its $i_{th}$ element is $PR(v_i)$ (PageRank of $v_i$). The notation $\vec{1}$ stands for a vector having all elements equal to $1$. The computation of PageRank values is done iteratively by defining a convergence stopping criterion $\epsilon$. At each computation step $t$, a new vector $(\vec{PR},t)$ is generated based on previous vector values $(\vec{PR},t-1)$. The algorithm stops computing values when the condition $|(\vec{PR},t)-(\vec{PR},t-1)|<\epsilon$ is satisfied.

\section{Security Correlation}
\subsection{Network Enrichment Using Passive DNS} \label{sec:pdns}
Passive DNS \cite{weimer2005passive}  replication is the process of capturing live DNS queries and/or their responses, and using this data to build partial replicas of as many DNS zones as possible. Passive DNS aims to make replication of the domain zones without the collaboration of zone administrators. A DNS sensor is used to capture the inter-server DNS communications.  Afterwards, the records of passive DNS are stored in a database where they can be queried. We can benefit from the passive DNS database in many ways. For instance, we can know the history of a domain name as well as the IP addresses it is/was pointing to. We can also find what domain names are hosted on a given name server or what domains are/(have been) pointing to a given IP address. There are a lot of use cases of passive DNS for security purposes (e.g., mapping criminal cyber-infrastructure \cite{antonakakis2010building}, tracking spam campaigns, tracking malware command and control systems, detection of fast-flux networks, security monitoring of a given cyber-infrastructure and botnet detection). In our context, we correlated \textsf{ToGather} with a passive DNS database ($30$Billion record) to enrich the investigation of Android malware by: (i) Finding suspicious domains that are pointing to a malicious IP address extracted from the analysis of a malware sample. (ii) Finding suspicious IP addresses that are resolved from a malicious domain that is extracted from the analysis of malware sample. (iii) Measuring the maliciousness magnitude of an IP address: a server identified by a malicious IP address that hosts many malicious activities. We could measure the maliciousness by counting the number of domains that resolve to this malicious IP address. Typically, these domains could be related to different malicious activities or a single one. (iv) Filtering outdated domain names: The passive DNS query generally returns timestamp information. \textsf{ToGather} could leverage the timestamps to filter out old domain names that are no longer active. 

\begin{figure}[!htb]
  \centering
      \includegraphics[width=0.45\textwidth]{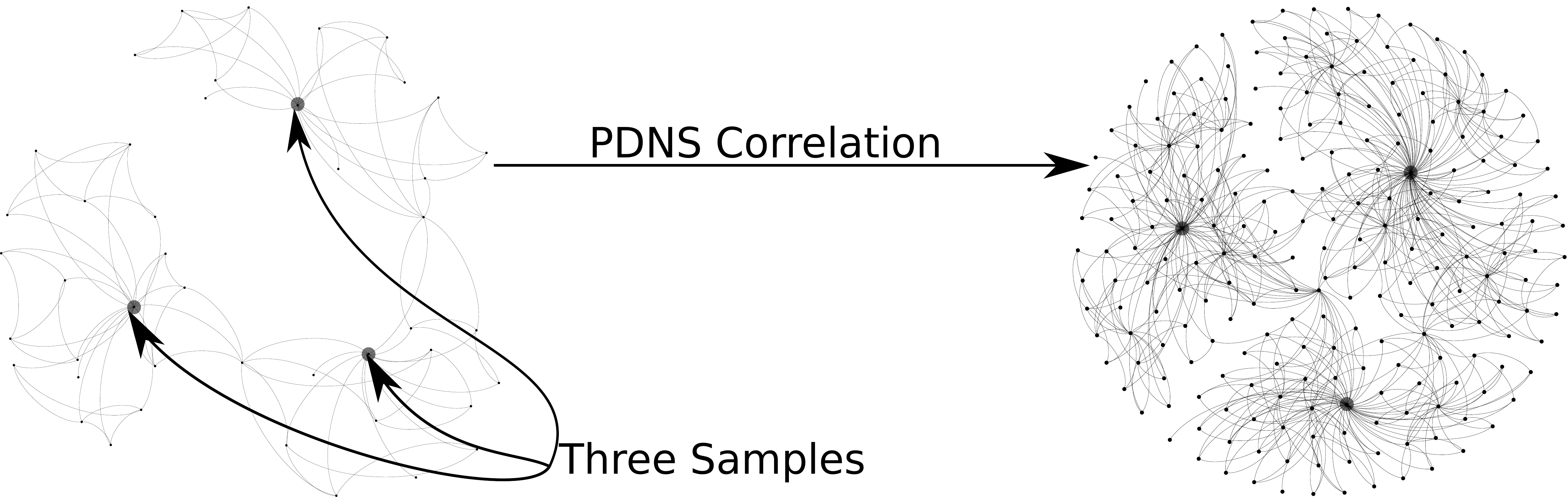}
  \caption{Threat Network With\&(out) Correlation}
  \label{fig:example_correlation}
\end{figure}

We consider passive DNS correlation as an optional component in \textsf{ToGather} workflow for two reasons. First, passive DNS could be missing to reproduce \textsf{ToGather} framework, since the security practitioner may not have access to such database. Second, the corpus of Android malware samples is enormous, and there are new feeds of malware samples every day. Hence, such large number of samples could fill the gap of passive DNS correlation due to the amount of the extracted network information. For example, as presented in Figure \ref{fig:example_correlation}, the threat network generated from three malware samples could be enhanced by the correlation with passive DNS.

\subsection{Threat Network Tagging} \label{sec:nettagging}
\textsf{ToGather} produces, from Android malware samples, a threat network that summarizes their malicious activities. Afterwards, \textsf{ToGather} detects and produces threat sub-networks if any. Besides, it helps prioritizing the actions to be taken to mitigate this threat using the PageRank algorithm. In this section, we go a step further towards the automatic investigation by leveraging other security feeds. Specifically, we aim at correlating threat networks with spam intelligence, reconnaissance intelligence, etc. The objective is to give a multi-dimensional view about the malicious activities that are related to the investigated Android malware family. Moreover, \textsf{ToGather} considers the correlation with network information from other platform malware; in the current setup, we correlate with PC malware from different operating systems.

\paragraph{PC Malware:} The adversary tends to have different malware samples in their arsenals to achieve their goal. Besides, different types of malware could be used to cover distinct platforms. The used malware samples run on many platforms, but they might share the elements of the same cyber-infrastructure run by the attacker. Therefore, finding other platform malware that share a similar threat network with a given Android malware sample, could help discovering other malware that is in the attackers' cyber-arsenals. Considering the previous case, \textsf{ToGather} tags every produced threat network by leveraging a database of network information extracted from PC malware VirusShare \cite{virusshare}. The malware database is continuously updated. The obtained information is identified by the malware hash and its malware family. The latter helps identifying PC malware (and their families) that share network information with the Android threat network.

\paragraph{Spam:} \textsf{ToGather} takes advantage of a spam database ($30$ Million record) to report the relationship between spamming campaigns and a given threat network. This information is precious for security analysts who are tracking spam campaigns.

\paragraph{Phishing:} Similarly to the spamming activity, we consider the phishing activity in \textsf{ToGather} tagging. Phishing activities aim at stealing sensitive information using fake web pages that are similar to known trusted ones. Typically, the attacker spread phishing sites using malicious URLs.  We extract only the domain name and store it in a phishing database ($5$ Million record).

\paragraph{Probing:} Probing \cite{panjwani2015experimental} is the activity of scanning networks over the Internet. The aim is to find vulnerable services. Probing is a significant concern in cyber-security because 50\% of cyber-attacks are preceded by network scanning activity \cite{panjwani2015experimental}. For this reason, \textsf{ToGather} considers tags of the threat network nodes if they are part of a probing activity. This pre-supposes the availability of a probing database ($300$ Million record) that contains IP addresses that have been part of scanning activities within the same epoch.  Probing could be derived from darknet traffic and the probing IP addresses could be persisted in a probing database.

\section{Experimental Results}

In this section, we present the evaluation results of our proposed system. The evaluation's goal is to assess the effectiveness of \textsf{ToGather} framework on giving a situational awareness from a set of Android malware samples. In our experimentation, we consider two cases of the entered malware samples: (a) The samples belong to the same Android malware family; here we look at the threat network of the given family and its sub ones. (b) The samples belong to different Android malware families; here we investigate the relation between the various families of Drebin dataset and how the threat network of the families could be distinguishable from other ones. Notice that the network information will be hidden in the result due the sensibility and confidentiality of this information (i.e., domains and IP addresses). Instead, we focus on the cyber-infrastructure of the malware samples, i.e., how the sub-threat networks could be apparent in the global threat network of Android malware family. Finally, we show the tagging result of the resulting threat network.

\subsection{Android Malawre Dataset}
In the evaluation,  we use a real Android malware dataset from Drebin \cite{arp2014drebin}, a known dataset that contains samples labeled with their families. Drebin dataset \cite{Drebin_Dataset} contains $5560$ labeled  malware samples from $179$ families \cite{Drebin_Dataset}., as shown in Table \ref{tab:dataset_family_number}. 

It is important to stress that Drebin contains all the samples of Genome dataset \cite{Android_Malware}. As a ground truth for the malware labeling, we take the label provided by Drebin since there are some differences between Genome and Drebin dataset labeling.  For example, Genome recognizes different versions of DroidKungFu malware (1, 2 and 4), where  Drebin has only DroidKungFu.

\begin{table}[!htb]
\centering
\begin{scriptsize}
\begin{tabular}{|l||l|c|}
\toprule
{} & Malawre Family & Number of Samples \\ \hline
\midrule 
0 &  FakeInstaller & 925 \\ \hline 
1 &    DroidKungFu & 667 \\ \hline
2 &      Plankton & 625 \\ \hline 
3 &      Opfake & 613 \\ \hline 
4 &      GinMaster & 339 \\ \hline 
5 &     BaseBridge & 330 \\ \hline 
6 &       Iconosys & 152 \\ \hline 
7 &        Kmin & 147 \\ \hline \bottomrule
\end{tabular}
\end{scriptsize}
\caption{Dataset Description By Malware Family} 
\label{tab:dataset_family_number}
\end{table}

\subsection{Implementation}
We have implemented \textsf{ToGather} using \textit{Python} programming language. In the static analysis, to perform reverse engineering of the \textit{Dex} byte-code, we use \textit{dexdump}, a tool provided with Android SDK. We extract the network information from the \textit{Dex} dis-assembly using regular expressions. Beside, \textsf{ToGather} extracts network information from static text content in the APK file of Android malware. 

\begin{figure}[!htb]
  \centering
      \includegraphics[width=0.45\textwidth]{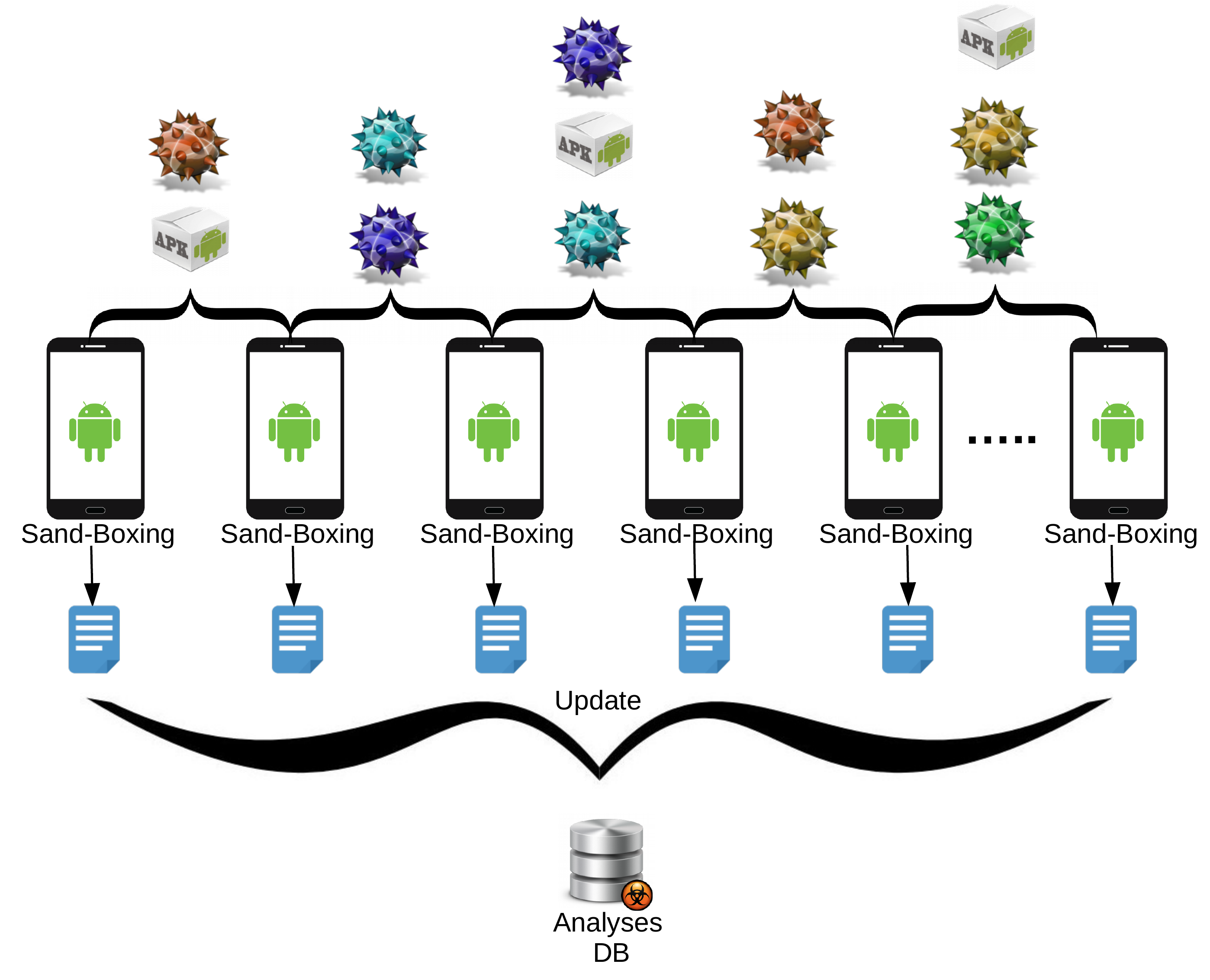}
  \caption{Sandboxing with Multi-Instance System}
  \label{fig:sandboxing}
\end{figure}

In the dynamic analysis, a cornerstone in \textsf{ToGather} framework is the sandboxing system, which heavily influences the produced analysis reports. We use \emph{DroidBox} \cite{droidbox_github}, a well-established sandboxing environment based on the Android software emulator \cite{android_emulator} provided by Google Android SDK \cite{android_sdk}. Running the app may not lead to a sufficient coverage of the executed app. As such, to simulate the user interaction with the apps, we leverage \textit{MonkeyRunner} \cite{monkeyrunner}, which produces random UI actions aiming for a broader execution coverage.  However, this makes the app execution non-deterministic since \textit{MonkeyRunner} generates random actions. Therefore, this yields different analysis reports for every execution, where the accuracy of the results may vary. To tackle this issue, we run the app in a sandboxing environment for a long time $T$ to assure the maximum of information in the resulting report. On the other hand, a long time $T$ could lead to execution bottleneck since \emph{DroidBox} can only handle one app at a time. In this context, executing the dataset apps in a sandboxing environment is a computation bottleneck in \textsf{ToGather}. To overcome this challenge, we develop a multi-worker sandboxing environment to exploit the maximum available resources and boost the sandboxing task. as depicted in Figure \ref{fig:sandboxing}.  Finally, since the generated reports are semi-structured (JSON files), we straightforwardly extract the network information from the specific fields.


\subsection{Drebin Threat Network} \label{sec:drebin_results}
\begin{figure}[!htb]
  \centering
      \includegraphics[width=0.33\textwidth]{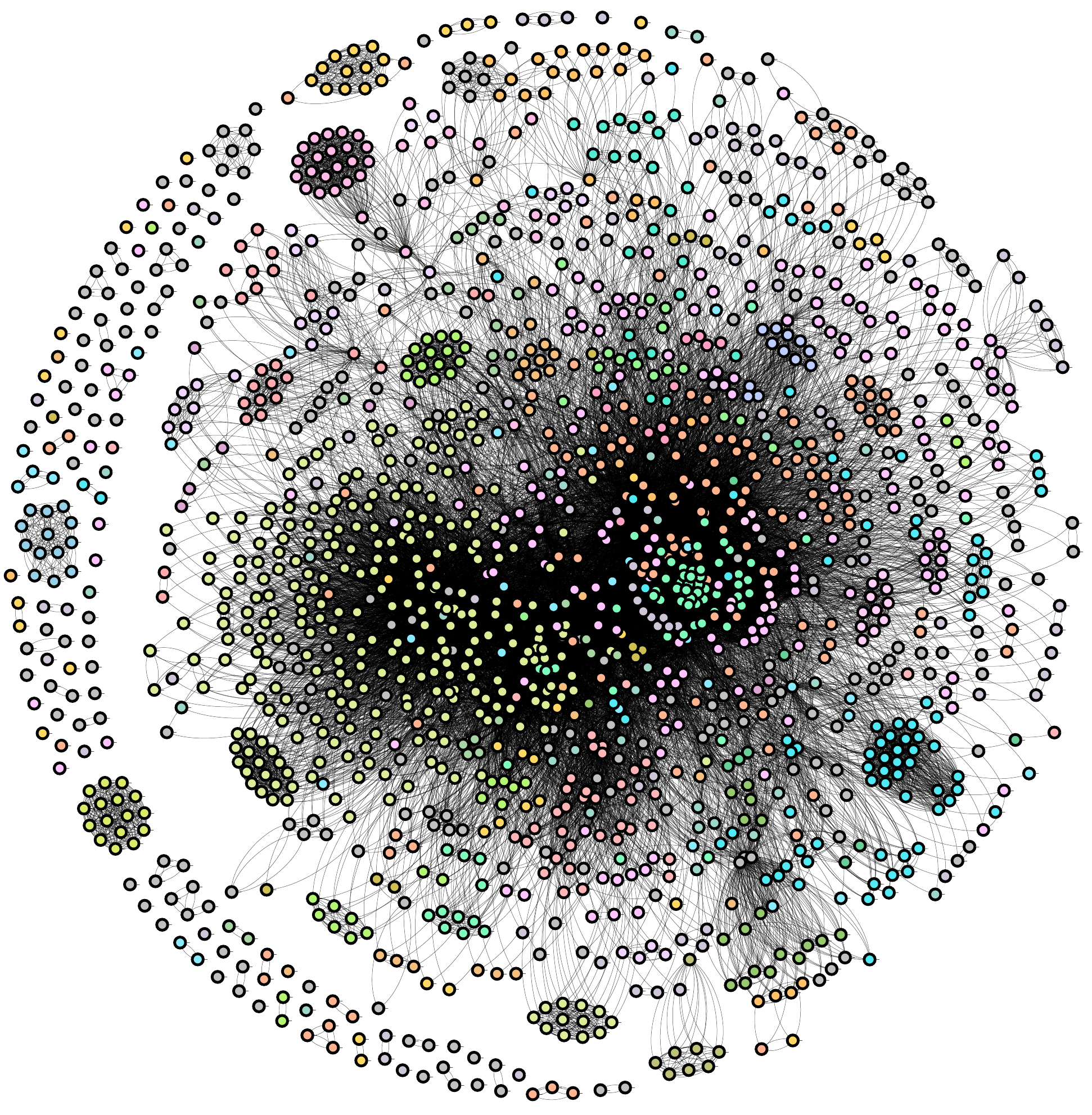}
  \caption{Network Information Drebin Dataset}
  \label{fig:allnet}
\end{figure}

\begin{figure}[!htb]
  \centering
      \includegraphics[width=0.33\textwidth]{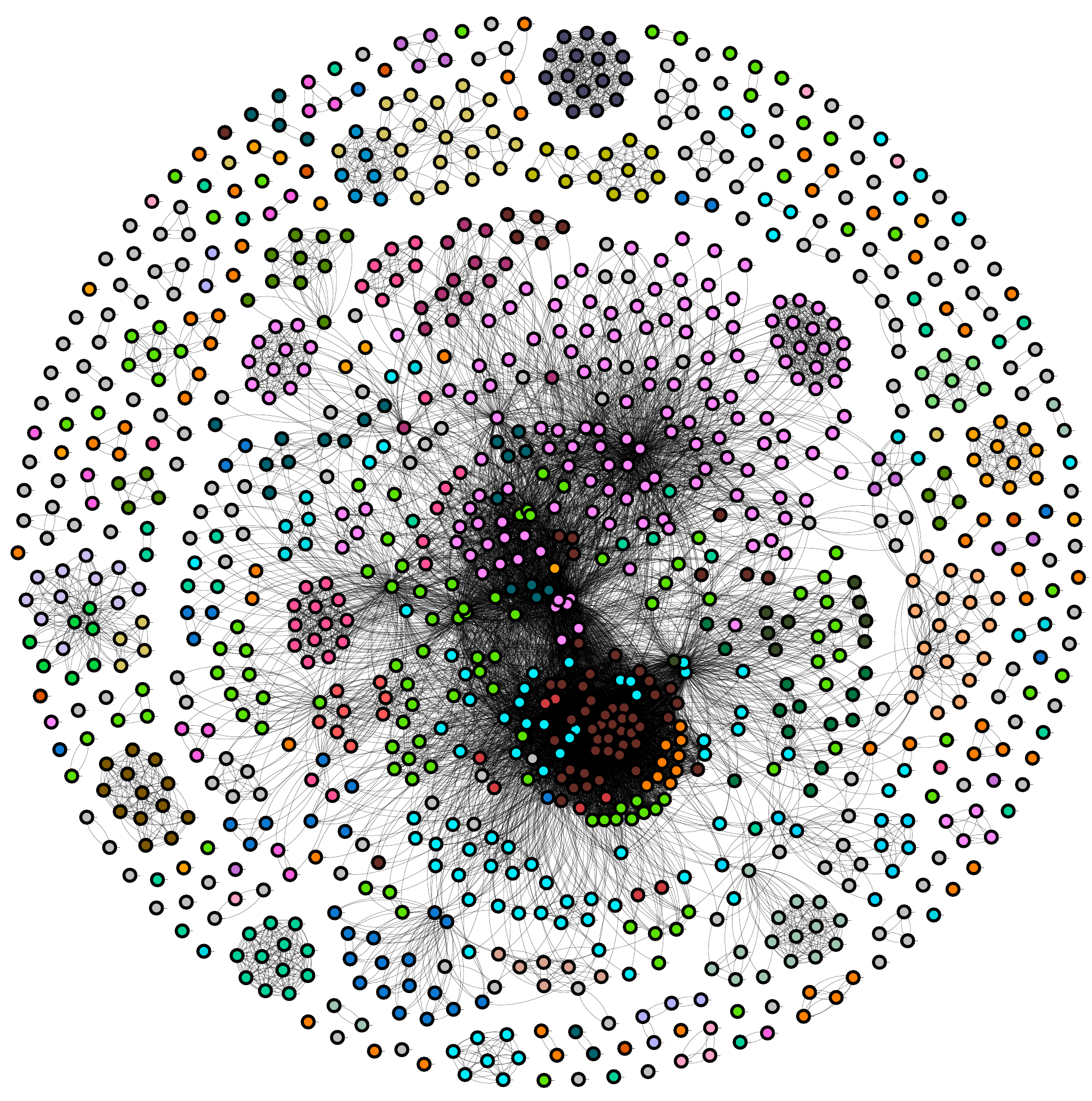}
  \caption{Domain Names Drebin Dataset}
  \label{fig:alldomainnet}
\end{figure}

In this section, we present the results of applying \textsf{ToGather} framework on the samples of Drebin dataset with all the $179$ families. Figure \ref{fig:allnet} depicts the threat network information (domain names and IP addresses) of Drebin dataset, where each family is represented by a different color. Although the threat network is noisy, we could visually distinguish some connected communities with the same nodes' color, i.e., the same malware family. This initial observation enhances the need for the community detection module in the \textsf{ToGather} framework. The community here is a set of graph nodes that are highly connected even though they share some links with external nodes. In Figure \ref{fig:alldomainnet}, we consider only the domain names; here we could distinguish more sub threat networks having nodes from the same malware family. We choose to filter all the IP addresses for Drebin dataset due to our observation during the experimentation process: (i) Some malware samples contain a significant malware number of IP addresses; exceeding, in some cases, $100$ IPs such as Plankton sample with MD5 hash \begin{scriptsize}\textbf{3f69836f64f956a5c00aa97bf1e04bf2}\end{scriptsize}. The adversary could aim to deceive the investigator by overwhelming the app with fake IP addresses along with used ones; this issue will be discussed in Section \ref{sec:futurework}. (ii) A big portion of the IP addresses are part of cloud companies infrastructure; we filter most of the public ones, but there are plenty of less known infrastructures in other countries. (iii) In most cases, the adversary utilizes domains for the malicious activity due to the low cost and the flexibility compared to IP addresses. In this experimentation, we consider only the domain names, but the security analyst could include the IP addresse when needed.

Using all Drebin dataset ($179$ malware families) to produce the Threat Network is an extream use case for \textsf{ToGather} framework; few malware families is a typical use case when we aim to investigate the threat networks relations. However, even with all Drebin dataset, \textsf{ToGather}, as presented in Figure \ref{fig:alldomainnet}, shows promising results, where we could see many sub-threat networks with(out) links to other nodes. By considering only domain names in Figure \ref{fig:alldomainnet}, it is noticeable that the size of the threat network significantly decreases by removing the IP addresses; normally there are significantly more domains than IP addresses in the Android apps. However, this is due to the extream whitelisting of domains compared to IPs (more than 1 million domain) and the size of Drebin dataset. At this stage, we do not present the community detection and page ranking on the threat network; this will be conducted on a one-family use case in the next section.

\textsf{ToGather} leverages different malicious datasets, as previously described in Section \ref{sec:nettagging}, to tag the nodes of the produced threat network. Figure \ref{fig:drebin_tagging} depicts the diverse malicious activities of the nodes from Drebin threat network. First, the table shows the top PC malware families which have shared network information with the Drebin threat network. For families' names, we adopt the \textit{Kaspersky} malware naming as our ground truth. Besides, Figure \ref{fig:drebin_tagging} shows the percentage of each malicious type in the Drebin threat network. The result shows that $56\%$ of the shared nodes have a spamming activity,  $40\%$ are related to PC malware, $3\%$ Scanning, and $1\%$ Phishing activities. Notice that the previous percentages are only from the shared nodes and not from all the threat network. Also, as we will discuss in Section  \ref{sec:futurework}, these results are not exhaustive because of the correlation datasets that obviously do not contain every malicious activity. We could extend the current correlation datasets to cover more suspicious activities in future work.

\begin{table}[!h]
\begin{minipage}{0.23\textwidth}
\includegraphics[width=1\textwidth]{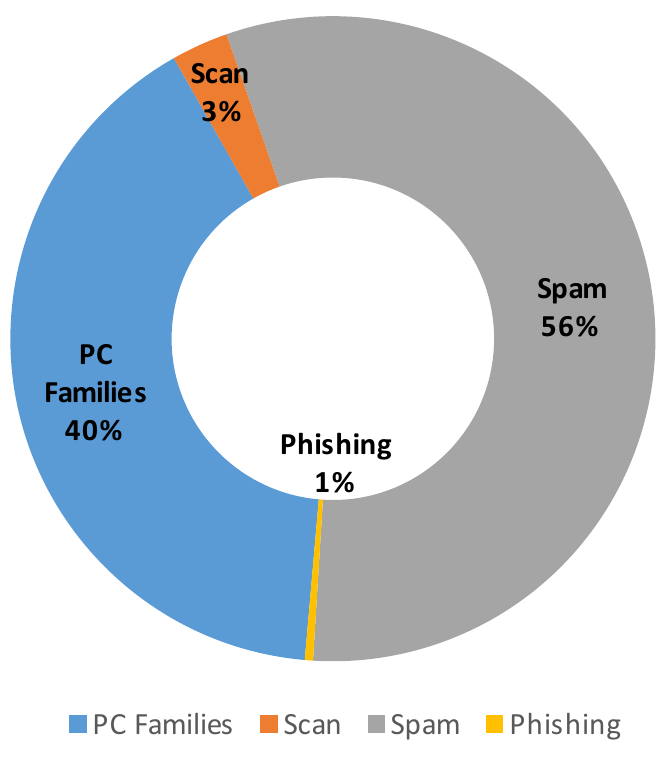}
\end{minipage}%
\hfill
\begin{minipage}{0.23\textwidth}
\centering
\begin{scriptsize}
\begin{tabular}{|c||c|c|}
    \hline \hline
    \#&  \textit{Family}     & \textit{Hits} \\ \hline\hline
	1 & Agent   \footnote{Kaspersky Naming}    & 1268 \\ \hline
	2 & VBNA        & 283 \\ \hline
	3 & Adload      &  152 \\ \hline
	4 & EgroupDial &  121 \\ \hline
	5 & TrustAsia  &  120 \\ \hline
	6 & Vobfus     &  88 \\ \hline
	7 & KuPlays    &  74 \\ \hline
	8 & Pipibo      &  72 \\ \hline
	9 & Sality       &  62 \\ \hline
\end{tabular} 
\end{scriptsize}
\end{minipage}%
\caption{ Drebin Dataset Tagging Results} 
\label{fig:drebin_tagging}
\end{table}

\subsection{Family Threat Networks} \label{sec:family_results}
In this section, we present the results of \textsf{ToGather} in its typical usage scenario where malware samples from the same family are analyzed. Figure \ref{fig:droidkungfunet} shows the steps of generating the threat networks from the DroidKungFu family sample.  First, \textsf{ToGather} produces the threat network including network information collected from the DroidKungFu analysis and Passive DNS correlation, as shown in Figure \ref{fig:droidkungfu_00}. Afterward, \textsf{ToGather} filters the whitelist network information. The results, as in Figure \ref{fig:droidkungfu_01}, depict bright separated sub-threat networks without applying the community detection algorithm. This could be an insightful result for the security practitioner, especially that this sub-threat network contains network information exclusively from some samples. \textsf{ToGather} goes a step ahead by applying both community detection (Resolution hyperparameter $r = 3$) and page ranking algorithms (damping factor $d = 0,85$ and stopping criterion $\epsilon = 0.001$ hyperparameters) to divide the network and rank the importance of the nodes respectively. The result is multiple sub-threat networks, with high interconnection and low intra-connection, representing the cyber-infrastructures of DroidKungFu malware family.

\begin{scriptsize}
\begin{figure}[ht!]
     \begin{center}
        \subfigure[Unfiltered]{%
            \label{fig:droidkungfu_00}
            \includegraphics[width=0.22\textwidth]{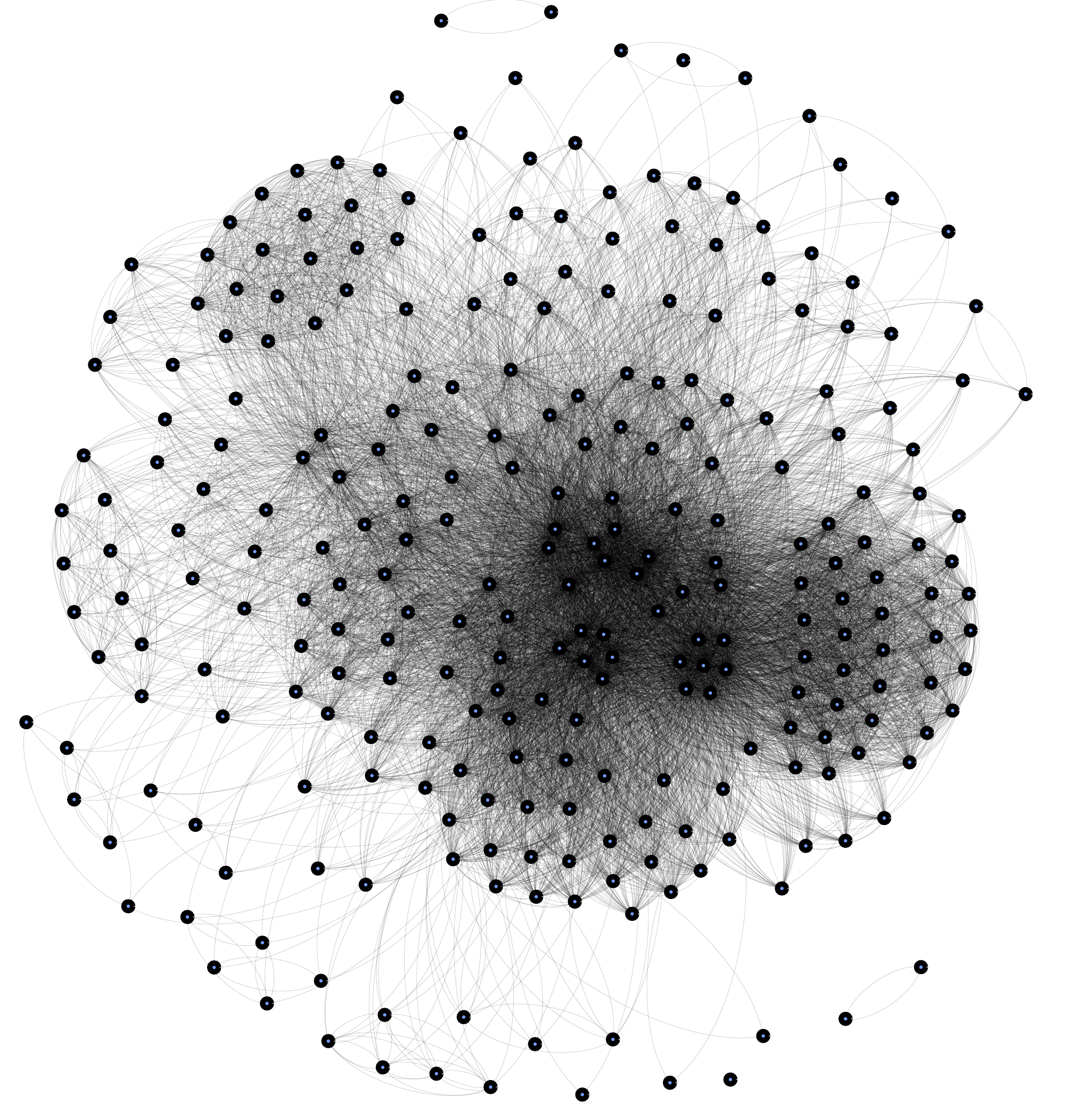}
        }
        \subfigure[Filtered]{%
           \label{fig:droidkungfu_01}
           \includegraphics[width=0.22\textwidth]{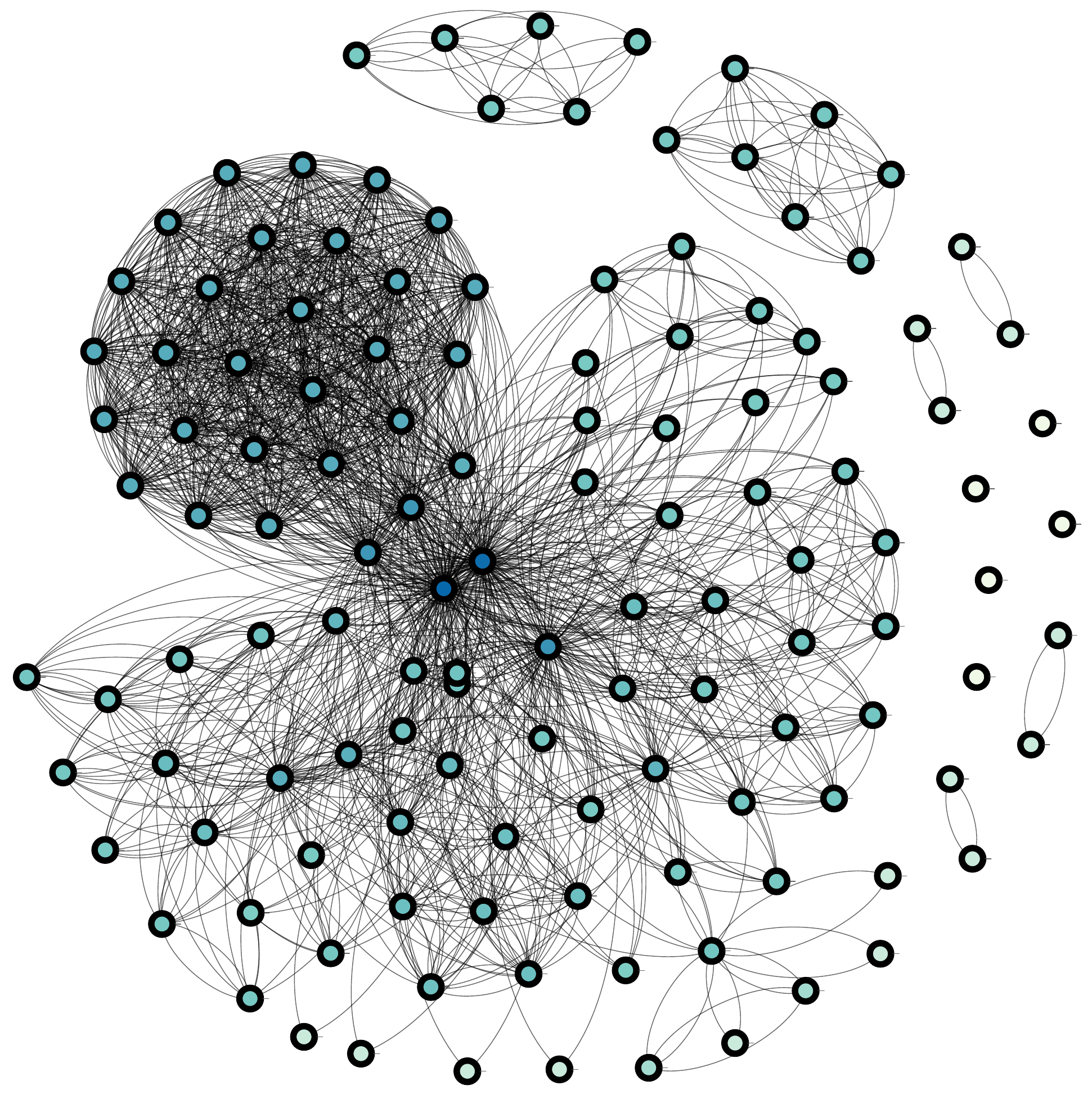}
        }
        \subfigure[Divide]{%
            \label{fig:droidkungfu_02}
            \includegraphics[width=0.22\textwidth]{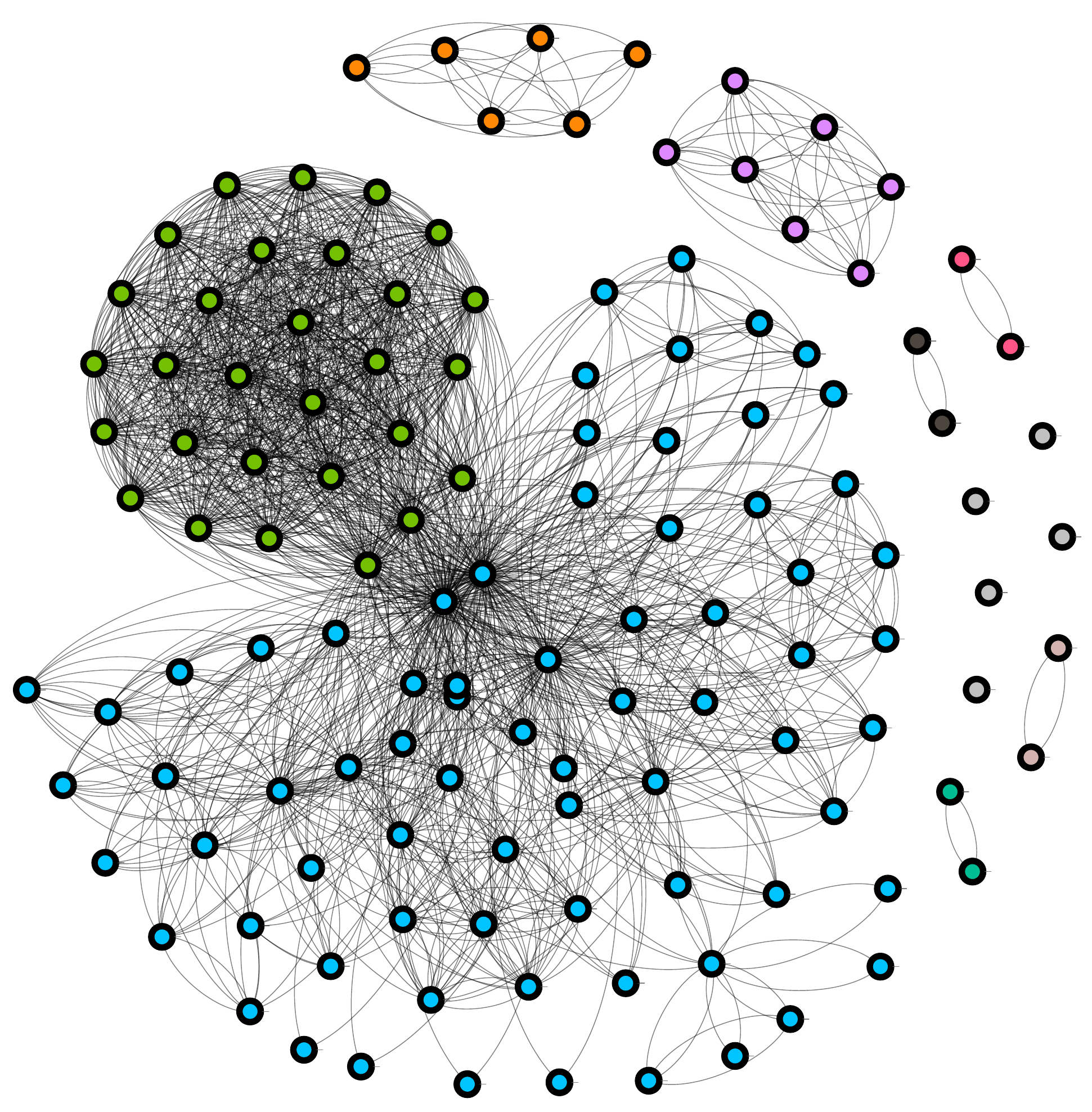}
        }
        \subfigure[Ranking]{%
           \label{fig:droidkungfu_03}
           \includegraphics[width=0.22\textwidth]{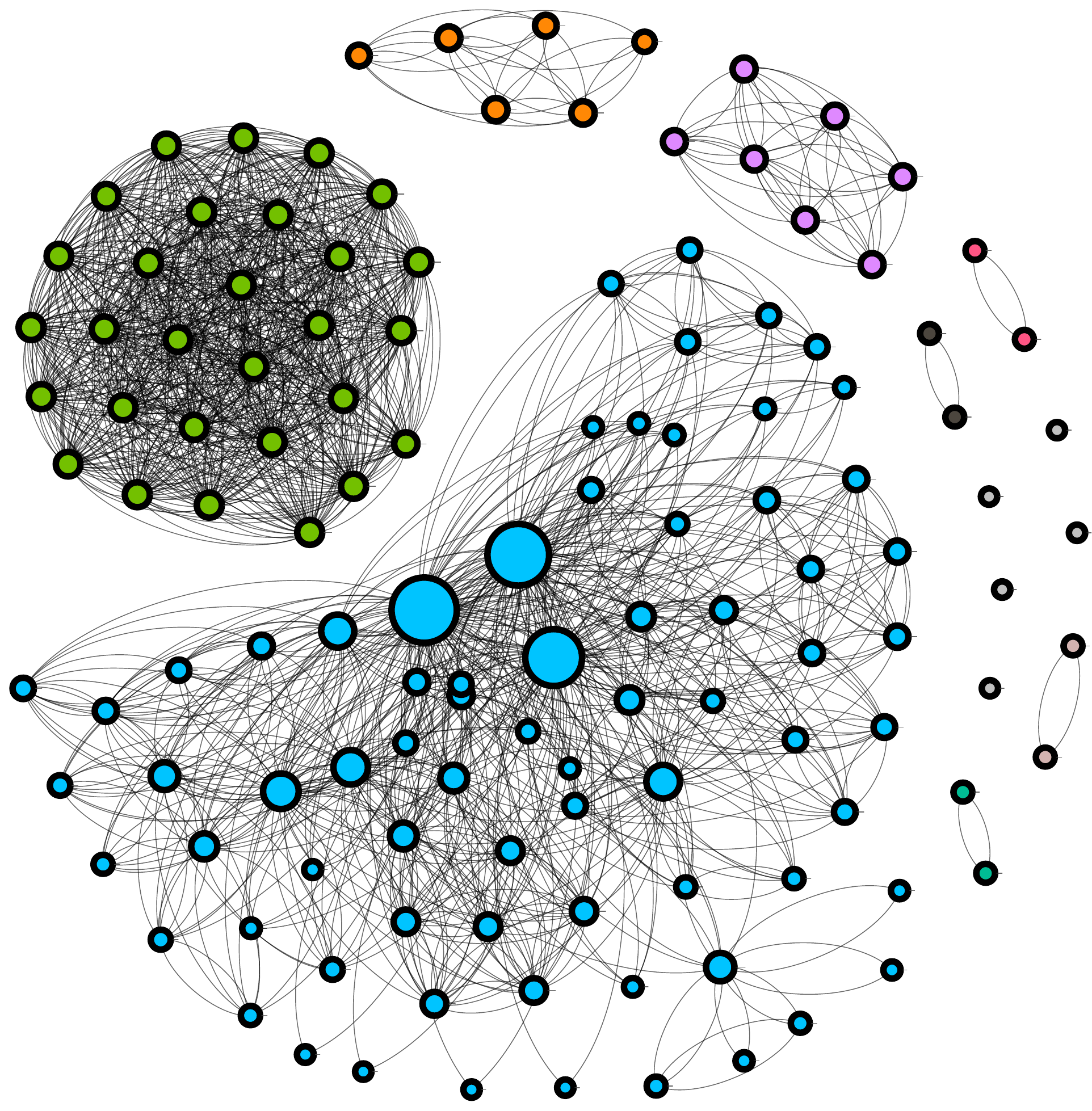}
        }
    \end{center}
    \caption{
        DroidKungFu Malware Threat Network 
     }
   \label{fig:droidkungfunet}
\end{figure}
\end{scriptsize}

Figure \ref{fig:basebrigdenet} shows \textsf{ToGather} results using Android malware samples from BaseBridge family. Similarly, after the filtering operation, we could easily distinguish small sub-threat networks. In same cases, the community detection task could be optional due to the clear separation between the sub-threat networks. For instance, Figure \ref{fig:drebinfamilynet} depicts the threat networks for GinMaster, Adrd, and Plankton Android malware families before and after the community detection task. Here, Adrd family clearly has multiple sub-threat networks without the need of the community detection function since it does not affect much the results. In the case of Plankton, it is necessary to detect and extract the sub-threat network.  

\begin{scriptsize}
\begin{figure}[ht!]
     \begin{center}
        \subfigure[Unfiltered]{%
            \label{fig:basebrigde_00}
            \includegraphics[width=0.14\textwidth]{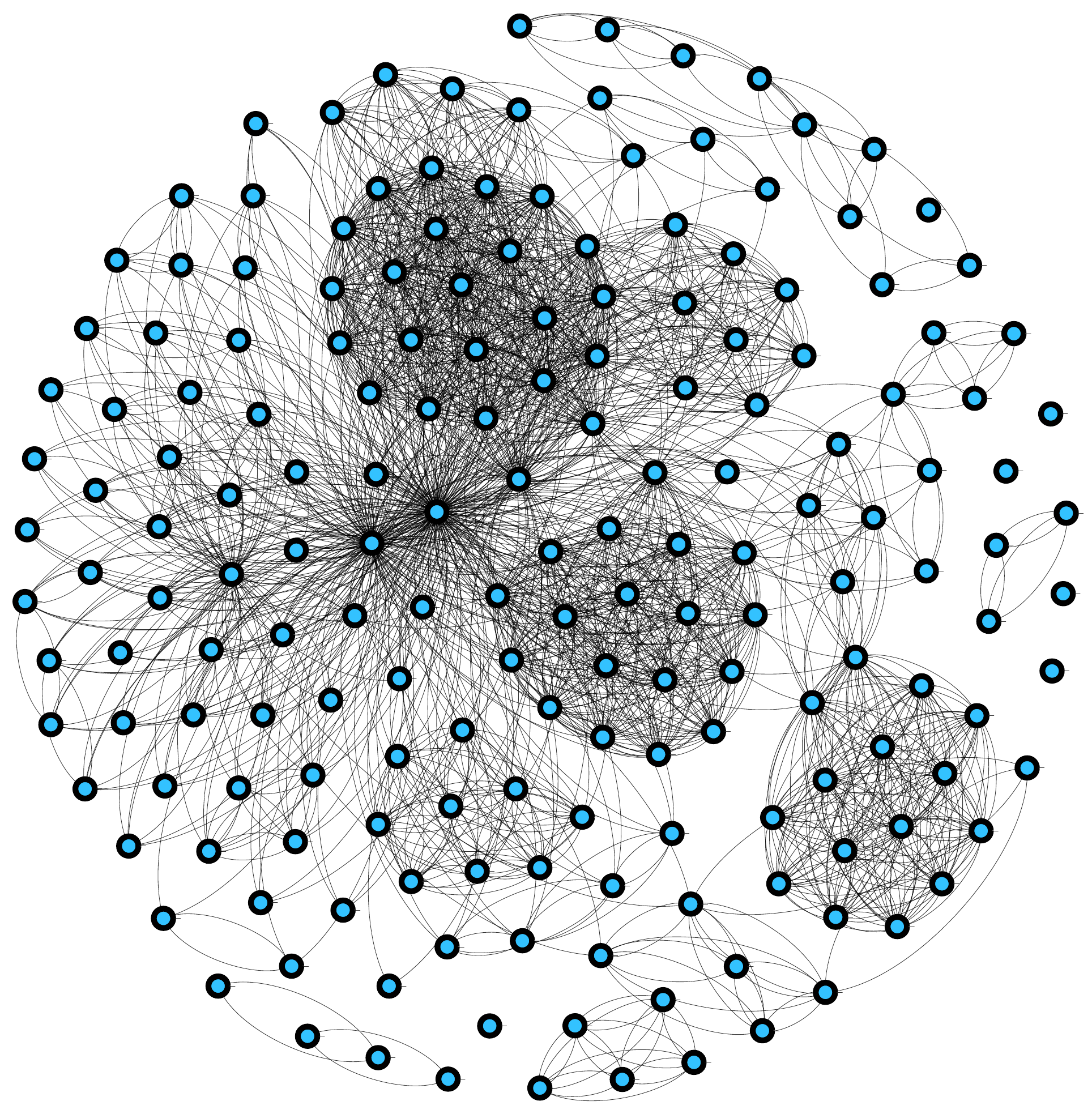}
        }
        \subfigure[Filter\&Divide]{%
           \label{fig:basebrigde_01}
           \includegraphics[width=0.14\textwidth]{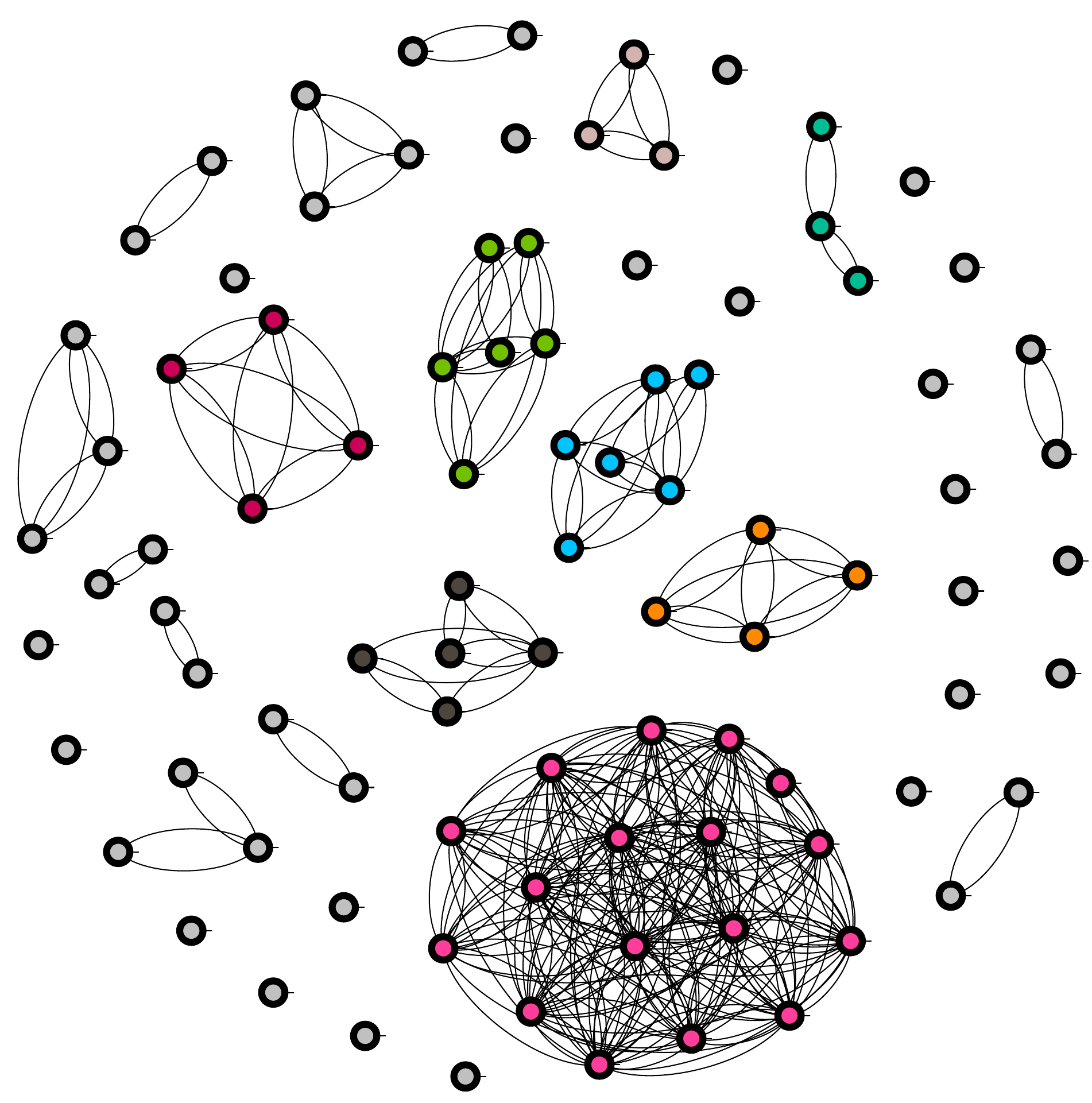}
        }
        \subfigure[Ranking]{%
            \label{fig:basebrigde_02}
            \includegraphics[width=0.14\textwidth]{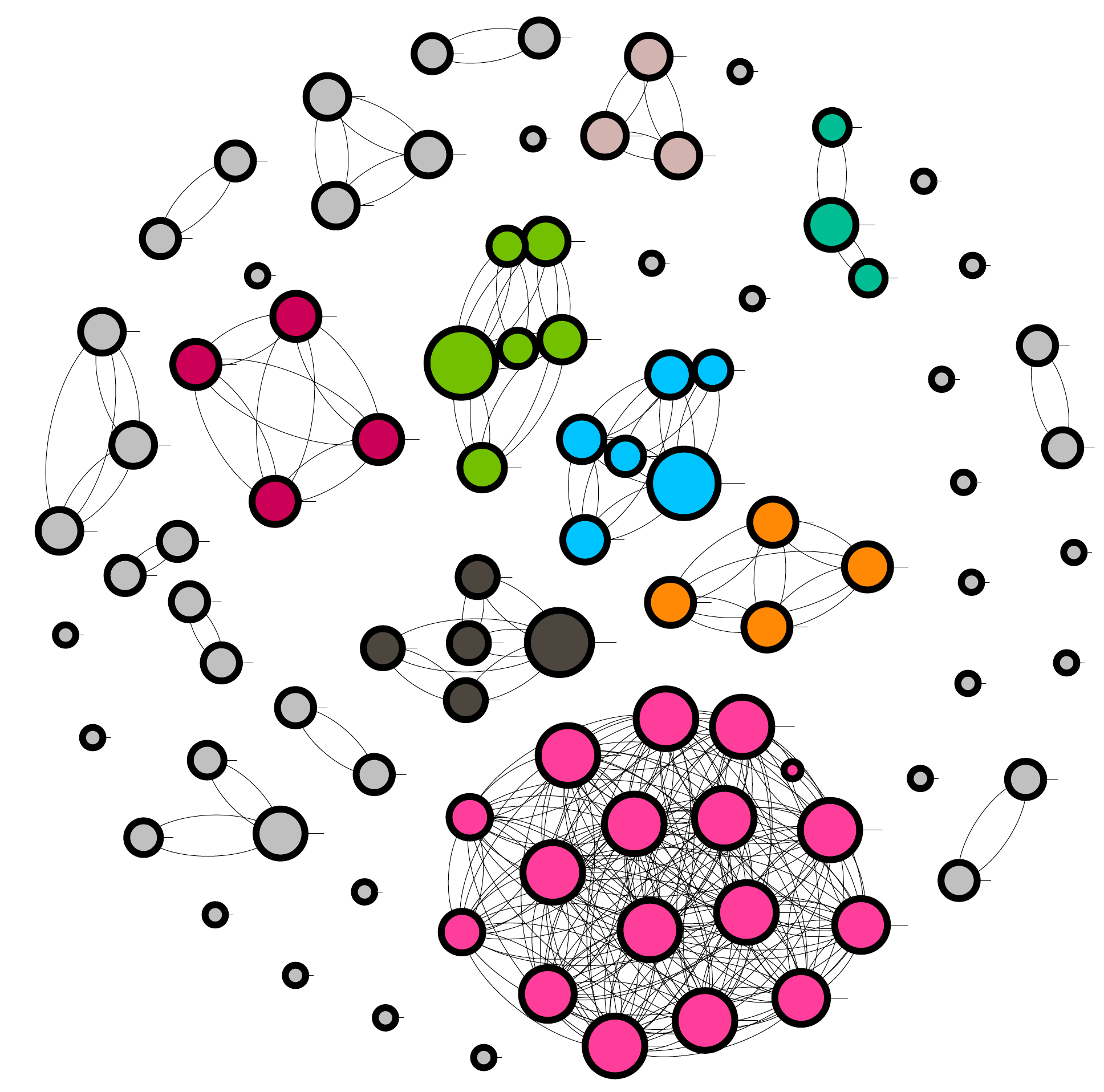}
        }
    \end{center}
    \caption{
        Basebrigde Malware Threat Network 
     }
   \label{fig:basebrigdenet}
\end{figure}
\end{scriptsize}

\begin{scriptsize}
\begin{figure}[ht!]
     \begin{center}
        \subfigure[Ginmaster (1)]{
            \label{fig:ginmaster_01}
            \includegraphics[width=0.14\textwidth]{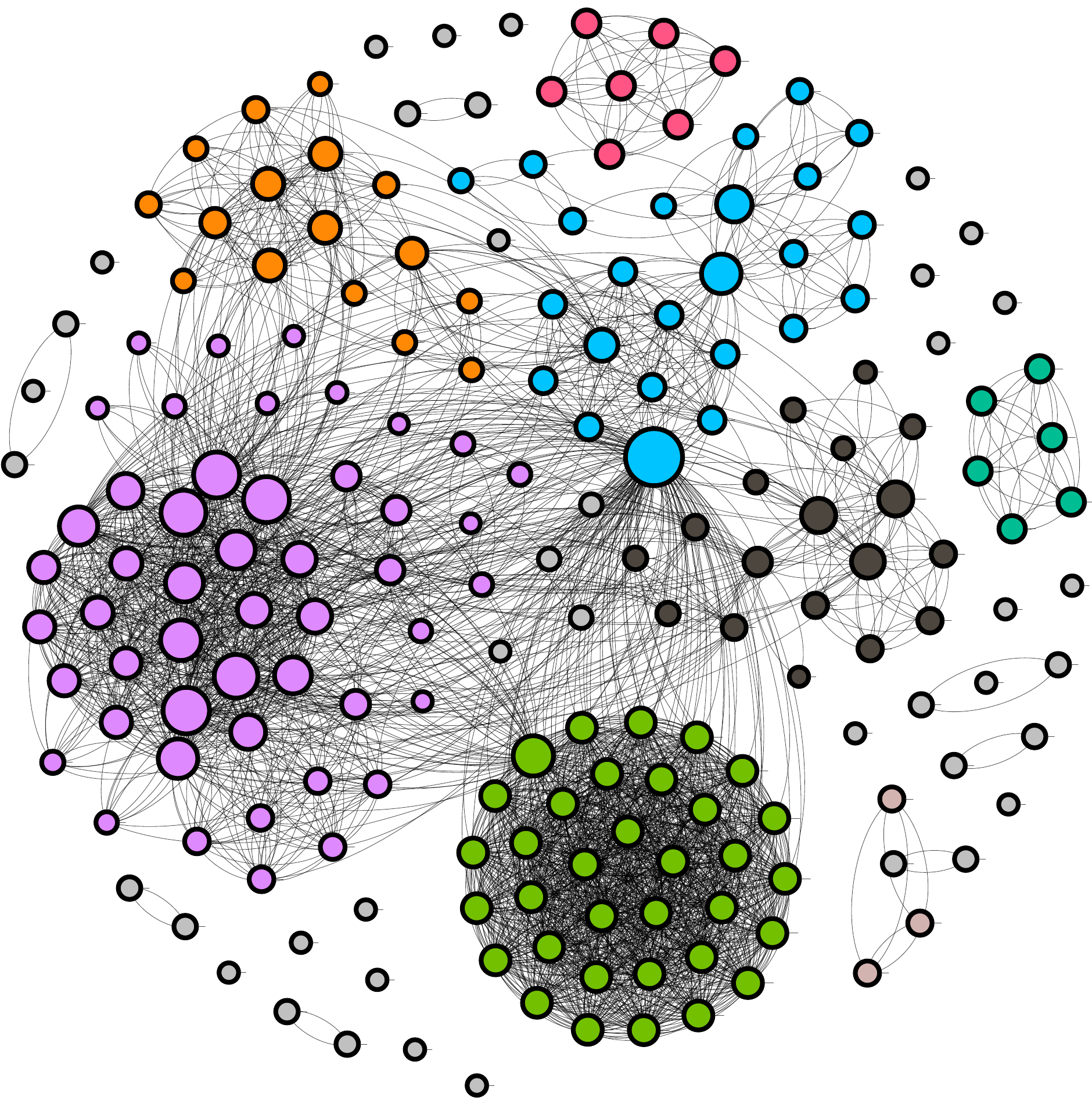}
        }
        \subfigure[Adrd (1)]{
           \label{fig:adrd_01}
           \includegraphics[width=0.14\textwidth]{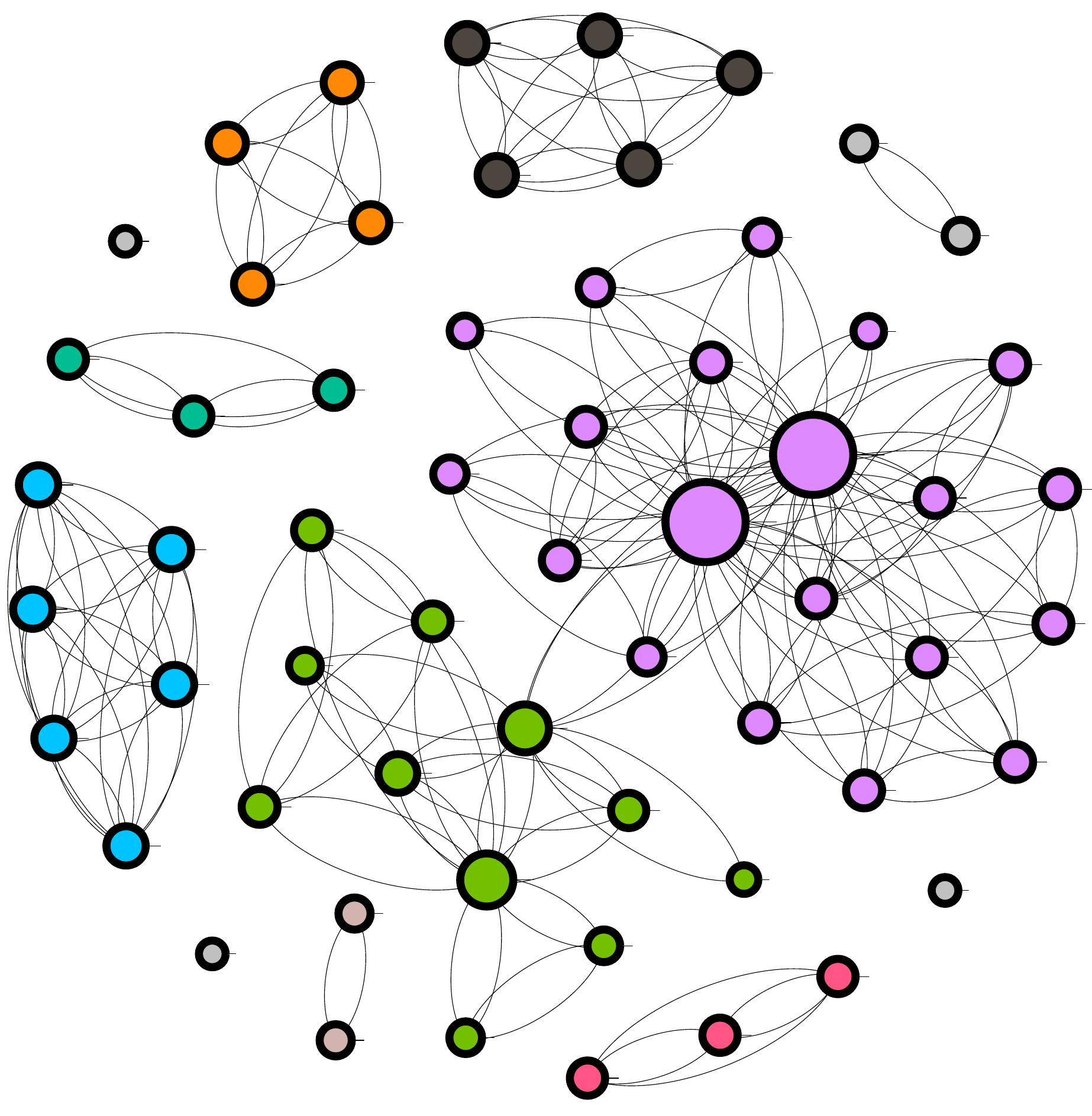}
        }
        \subfigure[Plankton (1)]{
            \label{fig:plankton_01}
            \includegraphics[width=0.14\textwidth]{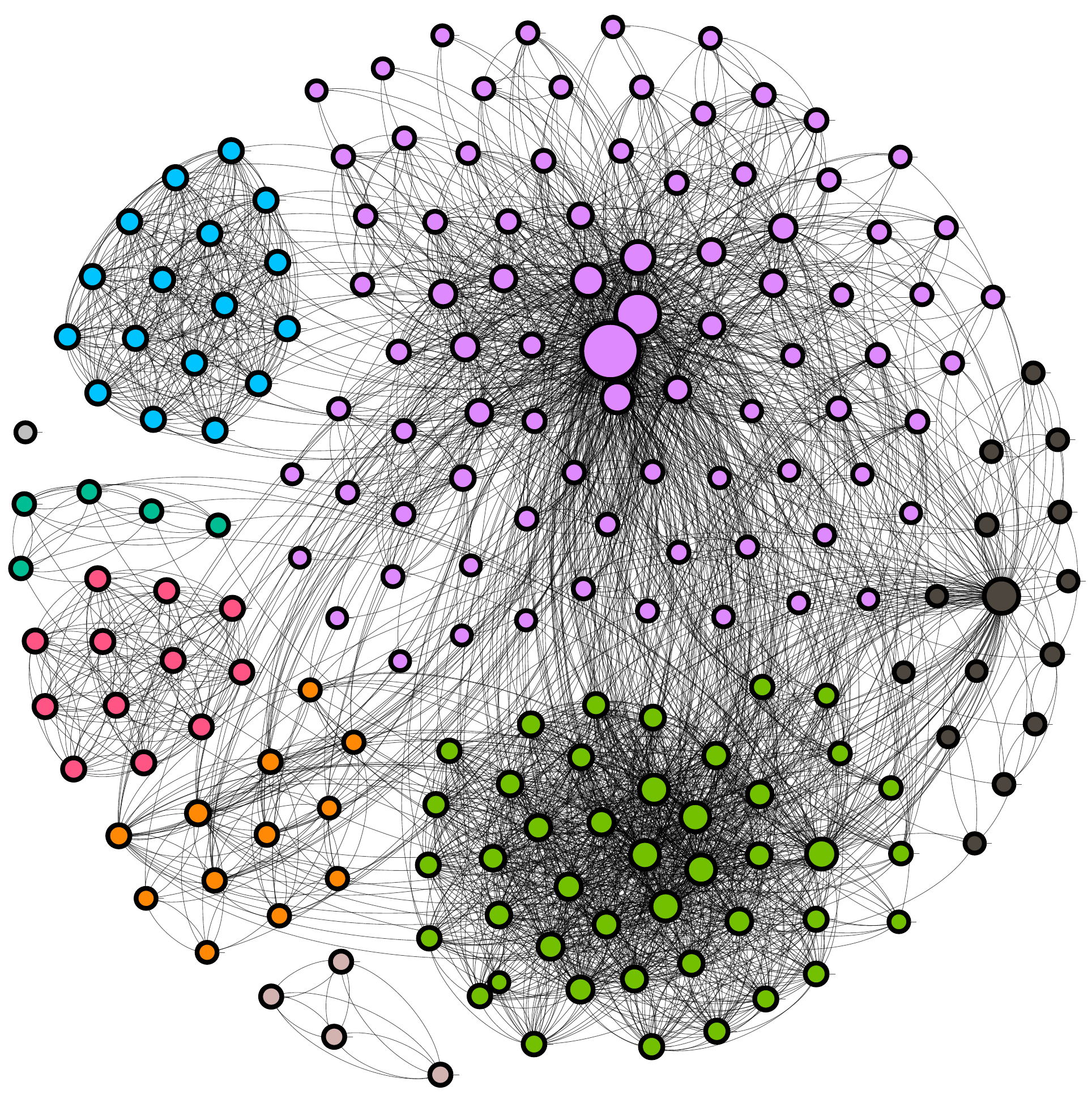}
        }
        \subfigure[Ginmaster (2)]{
            \label{fig:ginmaster_02}
            \includegraphics[width=0.14\textwidth]{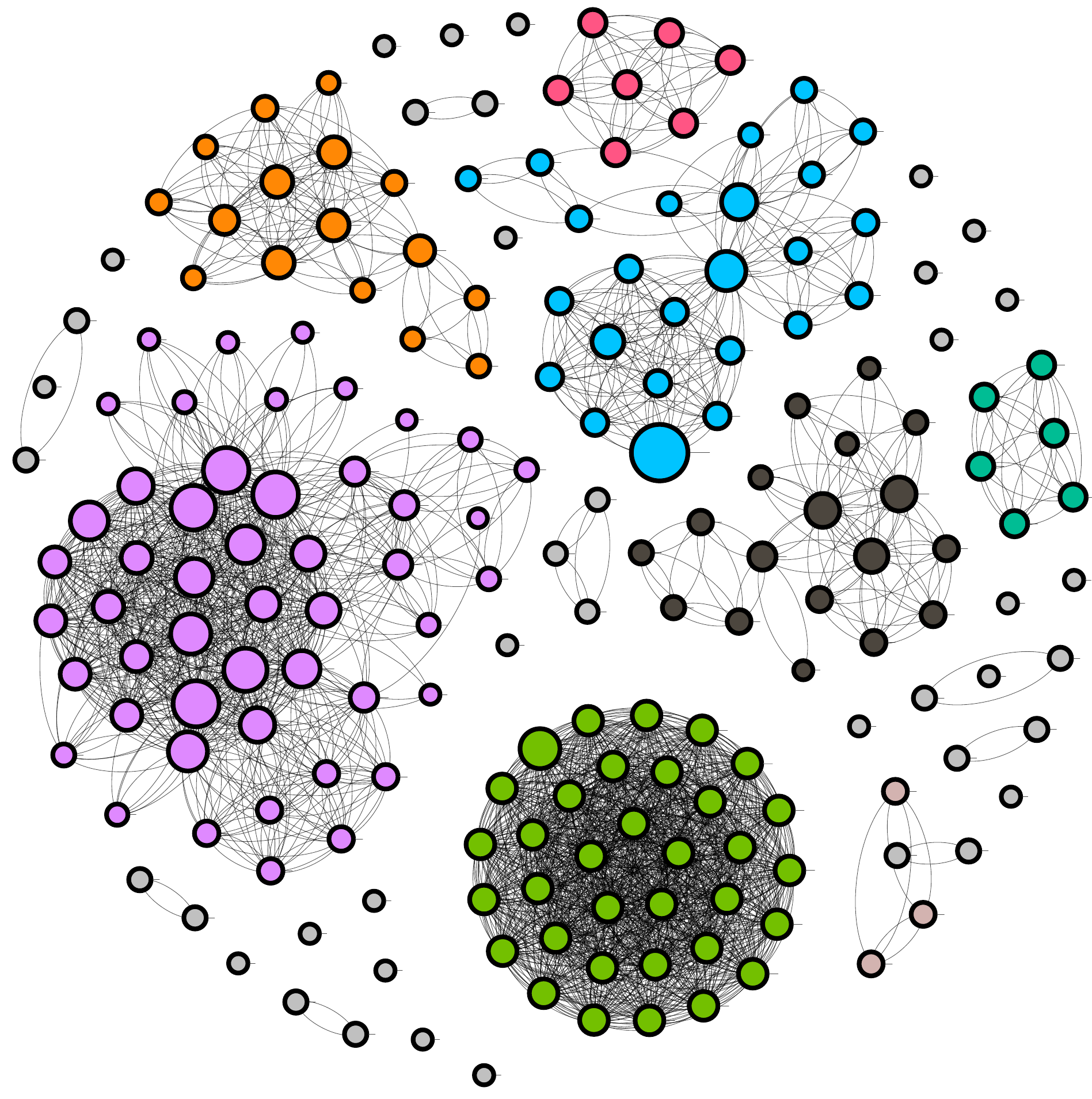}
        }
        \subfigure[Adrd (2)]{
            \label{fig:adrd_02}
            \includegraphics[width=0.14\textwidth]{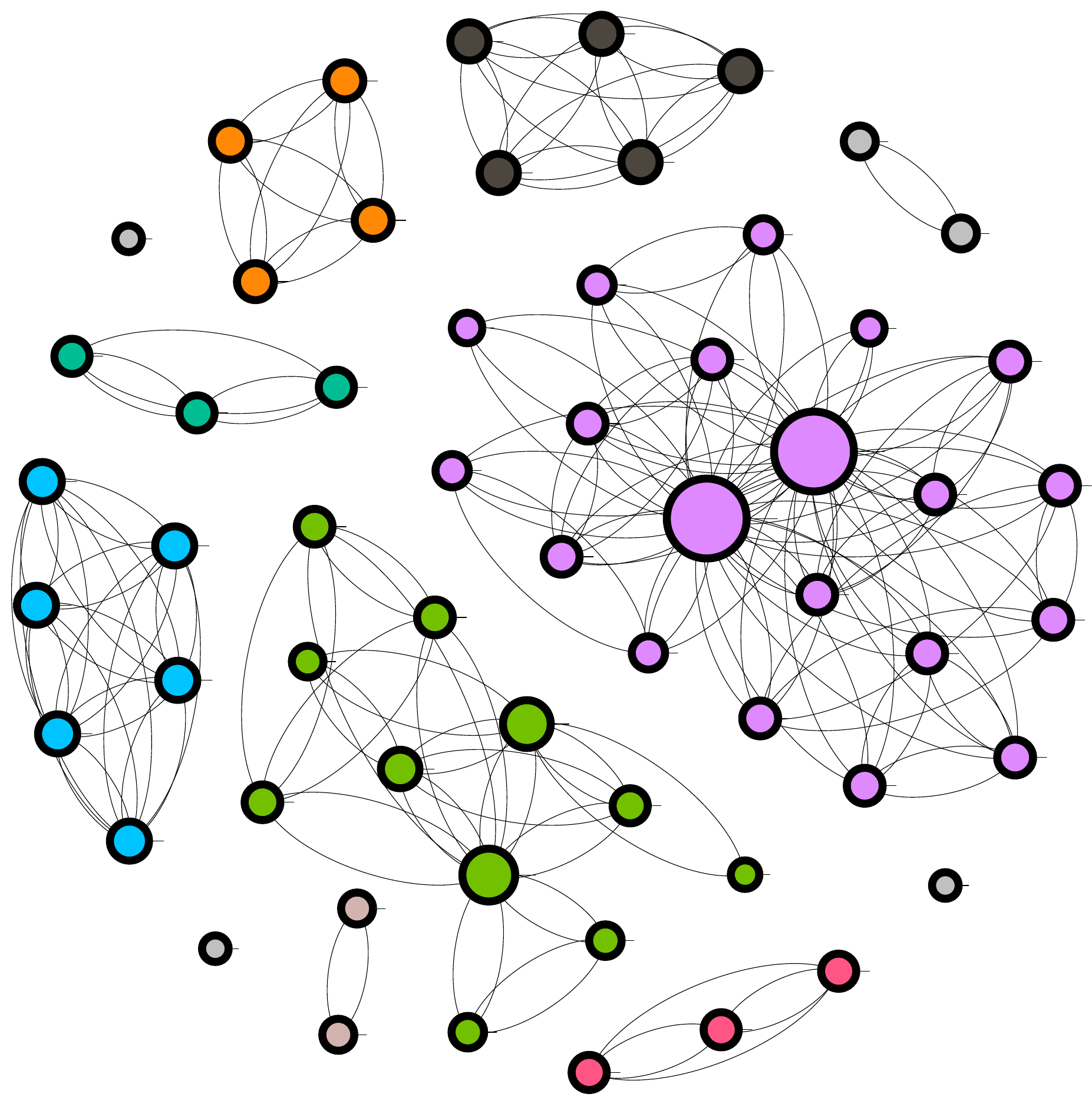}
        }
        \subfigure[Plankton (2)]{
            \label{fig:plankton_02}
            \includegraphics[width=0.14\textwidth]{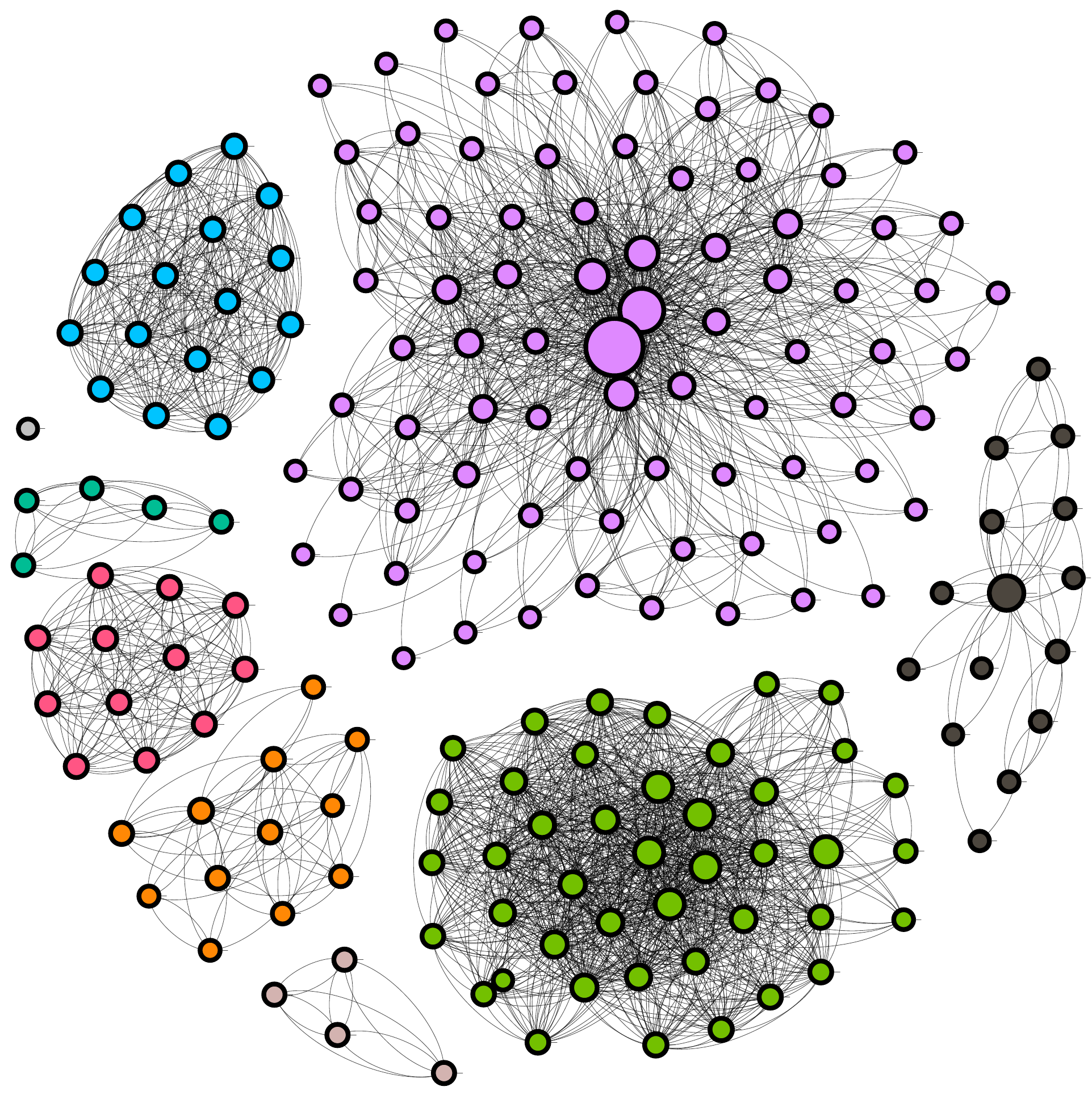}
        }
    \end{center}
    \caption{
        Android Families From Drebin Dataset  
     }
   \label{fig:drebinfamilynet}
\end{figure}
\end{scriptsize}

Tables \ref{tab:basemalware} and \ref{tab:kungfumalware} show the top PC malware families and samples that share the network information with  BaseBredge and DroidKungFu threat networks.  An important factor in the correlation is the explainability, where we could determine which network information is shared between the Android malware and the PC malware. This could help the security investigator to track the other dimension of the adversary cyber-infrastructure.

\begin{table}[!h]
\begin{minipage}{0.24\textwidth}
\centering
\begin{scriptsize}
\begin{tabular}{|c||c|c|c|}
    \hline \hline
     \#&  \textit{Sample}     & \textit{Hits}  \\ \hline\hline
	1 & ed7621ec4d \footnote{ MD5 Hash First 10 Chars}  &  2 \\ \hline
	2 & e3bc76d14c  & 2 \\ \hline
	3 & 503902c503  &  1 \\ \hline
	4 & bd9b87869b  &  1 \\ \hline
	5 & 8e0cf0a1ba6  &  1 \\ \hline
	6 & f8a5cac12dc &  1\\ \hline
	7 & 14db95e5f6  &  1 \\ \hline
	8 & 9b5b576ef3 &  1 \\ \hline
	9 & 2ec2abc28d &  1 \\ \hline
\end{tabular}
\end{scriptsize}
\end{minipage}%
\hfill
\begin{minipage}{0.23\textwidth}
\centering
\begin{scriptsize}
\begin{tabular}{|c||c|c|}
    \hline \hline
    \#&  \textit{Family}     & \textit{Hits} \\ \hline\hline
	1 & Agent   \footnote{Kaspersky Naming}    & 23 \\ \hline
	2 & Vobfus        & 21 \\ \hline
	3 & EgroupDial      &  13 \\ \hline
	4 & Badur  &  9 \\ \hline
	5 & LMN   &  7 \\ \hline
	6 & WBNA   &  4 \\ \hline
	7 & Pipibo    &  2 \\ \hline
	8 & Blocker  &  2 \\ \hline
	9 & Virut  &  2 \\ \hline
\end{tabular} 
\end{scriptsize}
\end{minipage}%
\caption{Top  PC Malware Related To BaseBridge Family } 
\label{tab:basemalware}
\end{table}

In addition to the PC malware tagging, we correlate with other cyber malicious activity datasets over the Internet.  Figure \ref{fig:malpies} presents the malicious activities of DroidKungFu and BaseBridge families that are related to their threat network. Here, we found that both families could be part of a spam campaign and have some scanning activity. Notice that these results represent a fraction of the actual activity because of the limited datasets. However, the previous fraction could be a good indicator for the security practitioner in the investigation process. 
 
\begin{table}[!h]
\begin{minipage}{0.24\textwidth}
\centering
\begin{scriptsize}
\begin{tabular}{|c||c|c|c|}
    \hline \hline
     \#&  \textit{Sample}     &  \textit{Hits}  \\ \hline\hline
	1 & 74529155cc  \footnote{ MD5 Hash First 10 Chars} &  3 \\ \hline
	2 & bd5a9f768cf &  2 \\ \hline
	3 & 259a244ab1 &  2 \\ \hline
	4 & 52da75225   &  1 \\ \hline
	5 & 11786afada  &   1 \\ \hline
	6 & ad5e6d577b &   1 \\ \hline
	7 & 9f4215bfc3  &   1 \\ \hline
	8 & 3c76ff67d0  &   1 \\ \hline
	9 & 117f21550   &   1 \\ \hline
\end{tabular}
\end{scriptsize}
\end{minipage}%
\hfill
\begin{minipage}{0.23\textwidth}
\begin{scriptsize}
\begin{tabular}{|c||c|c|}
    \hline \hline
    \#&  \textit{Family}     & \textit{Hits} \\ \hline\hline
	1 & Agent    \footnote{Kaspersky Naming}     & 33 \\ \hline
	2 & Adload        & 24 \\ \hline
	3 & TrustAsia      &  13 \\ \hline
	4 & KuPlays  &  11 \\ \hline
	5 & Pipibo   &  8 \\ \hline
	6 & FangPlay   &  5 \\ \hline
	7 & StartPage    &  4 \\ \hline
	8 & Injector  &  4 \\ \hline
	9 & Turbobit  &  4 \\ \hline
\end{tabular}
\end{scriptsize}
\end{minipage}%
\caption{Top  PC Malware Related To DroidKungFu Family } 
\label{tab:kungfumalware}
\end{table}

\begin{scriptsize}
\begin{figure}[ht!]
     \begin{center}
        \subfigure[BaseBridge]{%
            \label{fig:basebrigde_00}
            \includegraphics[width=0.22\textwidth]{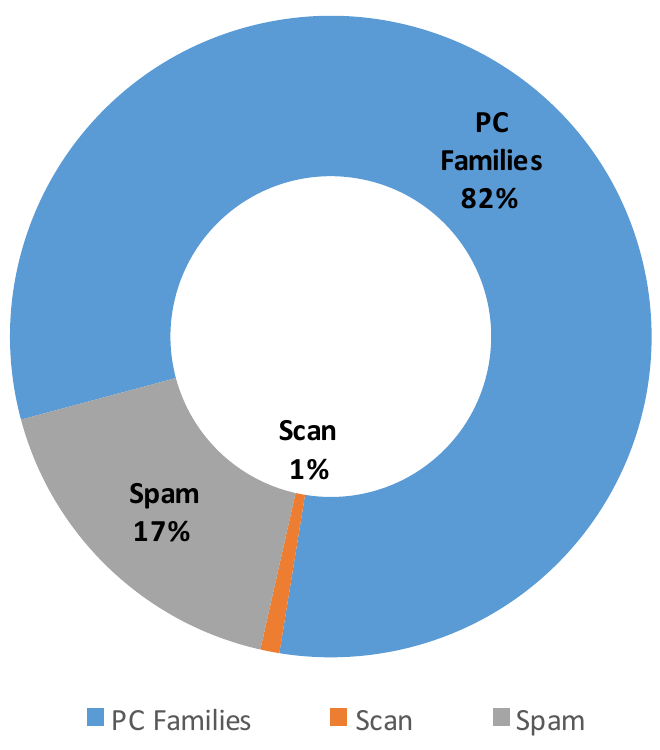}
        }
        \subfigure[DroidKungFu]{%
           \label{fig:basebrigde_01}
           \includegraphics[width=0.22\textwidth]{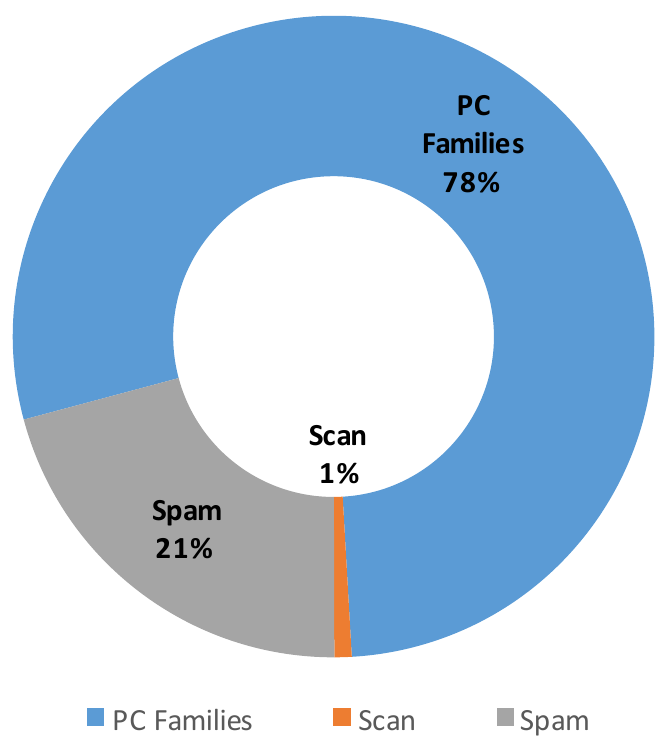}
        }
    \end{center}
    \caption{
        Maliciousness Tagging Per Family
     }
   \label{fig:malpies}
\end{figure}
\end{scriptsize}

\section{Discussion} \label{sec:discussion}
The results in the previous section show promising insights about the underlying cyber-infrastructures of the Android malware families. The produced threat networks could show one side of the adversary infrastructure, which is the Android malware one;  this side could lead to the complete threat network. Furthermore, all the previous results could be extracted automatically and periodically from a feed of Android malware samples belonging to one or various families. This requires fixing the hyperparameter related to the used algorithms, the community detection and the page ranking algorithms, as we did in our experimentation. Moreover, the number of the malware samples and their families could be a major hyperparameter that impacts the produced result. \textsf{ToGather} framework could tolerate having different Android malware families as presented previously, where we consider all the families of Drebin dataset ($179$ Families). Another important parameter is the whitelisting hyperparameter, which contains the number of domains from the top Alexa \& Quantcast lists. The latter could affect the result by introducing a lot of false positive domains. In our case, we consider the complete lists, which leads to  very few false positives. However, this could introduce false negatives by removing a site that is malicious. 

The concept of sub-threat network gives an insight on the possibility of having different threat networks, meaning multiple adversaries are reusing a family sample or one adversary uses distinct threat networks. Moreover, the sub-threat network helps the security analyst to tackle the cyber-threat sequentially, by focusing on one sub-threat network. Finally, passive DNS database has a paramount role in discovering the related domains and IPs without having all the samples of the malware family. Therefore, with a relatively  small set of samples, we could discover the threat network of the family. Finally, in the current implementation, we consider only the PC malware, spamming, phishing, and scanning, but the tagging could be extended to other security feeds.

\section{Limitation And Future Work} \label{sec:futurework}
\textsf{ToGather} results depend mainly on the input malware samples, which could affect the result in two ways. First, the adversary could include noisy network information in the actual Android package, thus  overwhelming the process of detecting the threat network. In the current \textsf{ToGather} setup, we consider both static and dynamic analyses to detect the network information. Afterward, we merge the result of each analysis to correlate with passive DNS and generate the threat network. This setup is venerable to such noise attack, but this could be mitigated by simply considering the intersection instead of the union of the network information analysis. Since the dynamic analysis result is more credible than the static one, the intersection of both analyses is much more credible. Also, the filtering operation could help mitigating such problem by removing possible noise that is part of the white list. To this end, \textsf{ToGather} adopts only static whitelisting, and we are planning to build a dynamic filtering system based on reputation similar to Notos \cite{antonakakis2010building}. Second, the Android malware family may rely on other means to connect the threat network by not using direct IPs or domains connection. Instead, the adversary could leverage legitimate services such as Twitter accounts to operate the malicious cyber-infrastructure. In this case, \textsf{ToGather} could not detect such threats since it relies mainly on network information. We consider such problem as future work, where we investigate different network information, including Twitter and IRC communication. Finally, \textsf{ToGather}'s current design produces a static threat network and sub-network; the time dimension is not provided. A workaround for this issue is to analyze the same malware family for different timestamps and keep the result of each timestamp. The latter could help in the study of the threat network over time. We plan to tackle this issue   more rigorously in future work.

\section{Related Work}
Previous works on Android malware mainly concentrate the actual malware sample using two basic approaches: static and dynamic analyses. Static analysis-based approaches \cite{arp2014drebin, karbab2016dna, feng2014apposcopy, karbab2016cypider, karbab2017android, karbab2018maldozer, yang2014apklancet, zhongyang2013droidalarm, sanz2014anomaly} 
rely on static features extracted from the app, such as requested permissions and APIs, to detect malicious apps. These methods are not resistant to obfuscation. On the other hand, dynamic analysis-based approaches \cite{canfora2016acquiring, karbab2017dysign, spreitzenbarth2013mobile, ali2016aspectdroid, zhang2013vetting, amos2013applying, wei2012android, wei2012android, huang2014asdroid} 
aim to identify a behavioral signature or anomaly of the running app. These methods are more resistant to obfuscation than the static ones as there are many ways to hide the malicious code, while it is difficult to hide the malicious behavior. The hybrid analysis methods \cite{yuan2014droid, grace2012riskranker, bhandari2015draco, vidas2014a5}
combine between static and dynamic analyses.  A significant number of papers has been recently proposed to detect repackaged apps by performing similarity analysis. The latter either identifies the apps that use the same malicious code (i.e., detection of malware families) \cite{kim2015structural, ali2015opseq, deshotels2014droidlegacy, zhou2012hey, gascon2013structural, suarez2014dendroid, kang2015detecting, lin2013identifying, faruki2015androsimilar}, or those that repackage the same original app (i.e., code reuse detection) \cite{chen2015finding, sun2015droideagle, zhou2012detecting, hanna2012juxtapp, crussell2012attack, zhou2013fast, crussell2013andarwin}. However, most of them consider only malware sample detection and not  the network dimension of the Android malware samples and their families. Differently, our work is novel in the sense that it represents the Android malware family by the underlying malicious cyber-infrastructure (i.e., threat network). The most similar work to our proposal is \cite{boukhtouta2015graph} \cite{nadji2013connected}, which studies the malicious threat networks in general. Our work is different from \cite{boukhtouta2015graph} \cite{nadji2013connected} by considering the Android malware sample as the seed to build the threat network. However, \cite{boukhtouta2015graph} \cite{nadji2013connected} deal with various sources as seeds. Besides, we propose \textsf{ToGather} as an online system to continuously generate the threat network of Android malware families in each epoch.

\section{Conclusion}
In this paper, we presented \textsf{ToGather} framework, a set of techniques, tools and mechanisms and security feeds bundled to automatically build a situational awareness about a given Android malware. \textsf{ToGather} leverages the stat-of-art of graph theory and multiple security feeds to produce insightful, granular, as well as actionable intelligence about malicious cyber-infrastructures related to the  Android malware samples. We experiment \textsf{ToGather} on real malware from Drebin Dataset\cite{arp2014drebin}. 


\bibliographystyle{plain}
\bibliography{references}

\begin{thebibliography}{10}

\bibitem{Drebin_Dataset}
{Drebin Dataset - http://tinyurl.com/pdsrtez}, 2015.

\bibitem{Android_Malware}
{Genome Dataset - http://tinyurl.com/combopx}, 2015.

\bibitem{alexatop_web}
{Alexa Top Sites - https://tinyurl.com/dkwndx}, 2016.

\bibitem{amazonip_web}
{Amazon IP Space - https://tinyurl.com/ze7pvdr}, 2016.

\bibitem{android_auto}
{Android Auto - http://tinyurl.com/hdsunht}, 2016.

\bibitem{android_emulator}
{Android Emulator - https://tinyurl.com/zlngucb}, 2016.

\bibitem{android_sdk}
{Android SDK - https://tinyurl.com/hn8qo9o}, 2016.

\bibitem{android_wear}
{Android Wear - http://tinyurl.com/qfa55o4}, 2016.

\bibitem{brillokey}
{Brillo - http://tinyurl.com/q5ko3zu}, 2016.

\bibitem{droidbox_github}
{DroidBox - https://tinyurl.com/jaruzgr}, 2016.

\bibitem{GData_2015_stats}
{G DATA Mobile Malware Report - http://tinyurl.com/jecm8gg}, 2016.

\bibitem{google_play}
{Google Play - https://play.google.com/}, 2016.

\bibitem{idc2016smartphone}
{Market Share - https://tinyurl.com/hy8z53j}, 2016.

\bibitem{monkeyrunner}
{MonkeyRunner - https://tinyurl.com/j6ruqkj}, 2016.

\bibitem{quantcast_web}
{Quantcast Sites - https://tinyurl.com/gmd577y}, 2016.

\bibitem{tracemyip_web}
{Tracemyip - Https://tinyurl.com/hpz3yfo}, 2016.

\bibitem{virusshare}
{Virusshare - http://tinyurl.com/j6vx542}, 2016.

\bibitem{ali2016aspectdroid}
Aisha Ali-Gombe, Irfan Ahmed, Golden~G {Richard III}, and Vassil Roussev.
\newblock {AspectDroid: Android App Analysis System}.
\newblock In {\em ACM Conf. Data Appl. Secur. Priv.}, 2016.

\bibitem{ali2015opseq}
Aisha Ali-Gombe, Irfan Ahmed, Golden~G {Richard III}, Vassil Roussev, Aisha~Ali
  Gombe, Irfan Ahmed, Golden G~Richard III, and Vassil Roussev.
\newblock {OpSeq: Android Malware Fingerprinting}.
\newblock In {\em Proc. 5th Progr. Prot. Reverse Eng. Work.}, 2015.

\bibitem{amos2013applying}
Brandon Amos, Hamilton Turner, and Jonathan White.
\newblock {Applying machine learning classifiers to dynamic android malware
  detection at scale}.
\newblock In {\em Wirel. Commun. Mob. Comput. Conf.}, 2013.

\bibitem{antonakakis2010building}
Manos Antonakakis, Roberto Perdisci, David Dagon, Wenke Lee, and Nick Feamster.
\newblock {Building a Dynamic Reputation System for DNS}.
\newblock {\em USENIX Secur.}, 2010.

\bibitem{arp2014drebin}
Daniel Arp, Michael Spreitzenbarth, Hubner Malte, Hugo Gascon, Konrad Rieck,
  Malte Hubner, Hugo Gascon, and Konrad Rieck.
\newblock {DREBIN: Effective and Explainable Detection of Android Malware in
  Your Pocket.}
\newblock In {\em Symp. Netw. Distrib. Syst. Secur.}, 2014.

\bibitem{bhandari2015draco}
Shweta Bhandari, Rishabh Gupta, Vijay Laxmi, Manoj~Singh Gaur, Akka Zemmari,
  and Maxim Anikeev.
\newblock {DRACO: DRoid analyst combo an android malware analysis framework}.
\newblock In {\em Int. Conf. Secur. Inf. Networks}, 2015.

\bibitem{fast08blondel}
Vincent~D. Blondel, Jean-Loup Guillaume, Renaud Lambiotte, and Etienne
  Lefebvre.
\newblock {Fast unfolding of communities in large networks}.
\newblock {\em J. Stat. Mech. Theory Exp.}, 2008.

\bibitem{borgatti2005centrality}
Stephen~P. Borgatti.
\newblock {Centrality and network flow}.
\newblock {\em Soc. Networks}, 27, 2005.

\bibitem{boukhtouta2015graph}
Amine Boukhtouta, Djedjiga Mouheb, Mourad Debbabi, Omar Alfandi, Farkhund
  Iqbal, May {El Barachi}, and May~El Barachi.
\newblock {Graph-theoretic characterization of cyber-threat infrastructures}.
\newblock {\em Digit. Investig.}, 2015.

\bibitem{brin1998anatomy}
Sergey Brin and Lawrence Page.
\newblock {The anatomy of a large-scale hypertextual Web search engine}.
\newblock {\em Comput. Networks ISDN Syst.}, 1998.

\bibitem{canfora2016acquiring}
Gerardo Canfora, Eric Medvet, Francesco Mercaldo, and Corrado~Aaron Visaggio.
\newblock {Acquiring and Analyzing App Metrics for Effective Mobile Malware
  Detection}.
\newblock In {\em 2016 ACM Int. Work. Secur. Priv. Anal.}, IWSPA '16, 2016.

\bibitem{chen2015finding}
Kai Chen, Peng Wang, Yeonjoon Lee, XiaoFeng Wang, Nan Zhang, Heqing Huang, Wei
  Zou, and Peng Liu.
\newblock {Finding Unknown Malice in 10 Seconds: Mass Vetting for New Threats
  at the Google-Play Scale}.
\newblock In {\em USENIX Secur. Symp.}, 2015.

\bibitem{crussell2012attack}
Jonathan Crussell, Clint Gibler, and Hao Chen.
\newblock {Attack of the clones: Detecting cloned applications on android
  markets}.
\newblock In {\em Eur. Symp. Res. Comput. Secur. Proc.} 2012.

\bibitem{crussell2013andarwin}
Jonathan Crussell, Clint Gibler, and Hao Chen.
\newblock {AnDarwin: Scalable Detection of Semantically Similar Android
  Applications}.
\newblock In {\em Eur. Symp. Res. Comput. Secur. Egham}, 2013.

\bibitem{deshotels2014droidlegacy}
Luke Deshotels, Vivek Notani, and Arun Lakhotia.
\newblock {Droidlegacy: Automated familial classification of android malware}.
\newblock In {\em Proc. ACM SIGPLAN Progr. Prot. Reverse Eng. Work.}, 2014.

\bibitem{faruki2015androsimilar}
Parvez Faruki, Vijay Laxmi, Ammar Bharmal, M.~S. Gaur, and Vijay Ganmoor.
\newblock {AndroSimilar: Robust signature for detecting variants of Android
  malware}.
\newblock 2015.

\bibitem{feng2014apposcopy}
Yu~Feng, Saswat Anand, Isil Dillig, and Alex Aiken.
\newblock {Apposcopy: Semantics-based detection of android malware through
  static analysis}.
\newblock In {\em ACM SIGSOFT Int. Symp. Found. Softw. Eng.}, 2014.

\bibitem{gascon2013structural}
Hugo Gascon, Fabian Yamaguchi, Daniel Arp, and Konrad Rieck.
\newblock {Structural detection of android malware using embedded call graphs}.
\newblock In {\em ACM Work. Artif. Intell. Secur.}, 2013.

\bibitem{grace2012riskranker}
Michael Grace, Yajin Zhou, Qiang Zhang, Shihong Zou, and Xuxian Jiang.
\newblock {Riskranker: scalable and accurate zero-day android malware
  detection}.
\newblock In {\em Proc. 10th Int. Conf. Mob. Syst. Appl. Serv.}, 2012.

\bibitem{hanna2012juxtapp}
Steve Hanna, Ling Huang, Edward~XueJun Wu, Saung Li, Charles Chen, and Dawn
  Song.
\newblock {Juxtapp: A scalable system for detecting code reuse among android
  applications}.
\newblock In {\em Detect. Intrusions Malware, Vulnerability Assess.} Springer,
  2012.

\bibitem{huang2014asdroid}
Jianjun Huang, Xiangyu Zhang, Lin Tan, Peng Wang, and Bin Liang.
\newblock {AsDroid: Detecting stealthy behaviors in android applications by
  user interface and program behavior contradiction}.
\newblock In {\em Proc. 36th Int. Conf. Softw. Eng.}, 2014.

\bibitem{kang2015detecting}
Hyunjae Kang, Jae-wook Jang, Aziz Mohaisen, and Huy~Kang Kim.
\newblock {Detecting and classifying android malware using static analysis
  along with creator information}.
\newblock {\em Int. J. Distrib. Sens. Networks}, 2015.

\bibitem{karbab2017dysign}
ElMouatez~Billah Karbab, Mourad Debbabi, Saed Alrabaee, and Djedjiga Mouheb.
\newblock {DySign: Dynamic fingerprinting for the automatic detection of
  android malware}.
\newblock {\em 2016 11th International Conference on Malicious and Unwanted
  Software, MALWARE 2016}, pages 139--146, 2017.

\bibitem{karbab2016cypider}
ElMouatez~Billah Karbab, Mourad Debbabi, Abdelouahid Derhab, and Djedjiga
  Mouheb.
\newblock {Cypider: Building Community-Based Cyber-Defense Infrastructure for
  Android Malware Detection}.
\newblock In {\em ACM Computer Security Applications Conference (ACSAC)},
  volume 5-9-Decemb, pages 348--362. ACM, 2016.

\bibitem{karbab2017android}
ElMouatez~Billah Karbab, Mourad Debbabi, Abdelouahid Derhab, and Djedjiga
  Mouheb.
\newblock {Android Malware Detection using Deep Learning on API Method
  Sequences}.
\newblock 2017.

\bibitem{karbab2018maldozer}
ElMouatez~Billah Karbab, Mourad Debbabi, Abdelouahid Derhab, and Djedjiga
  Mouheb.
\newblock {MalDozer: Automatic framework for android malware detection using
  deep learning}.
\newblock {\em Digital Investigation}, 2018.

\bibitem{karbab2016dna}
ElMouatez~Billah Karbab, Mourad Debbabi, and Djedjiga Mouheb.
\newblock {Fingerprinting Android packaging: Generating DNAs for malware
  detection}.
\newblock {\em Digit. Investig.}, 2016.

\bibitem{kim2015structural}
Junhyoung Kim, Tae~Guen Kim, and Eul~Gyu Im.
\newblock {Structural information based malicious app similarity calculation
  and clustering}.
\newblock In {\em Conf. Res. Adapt. Converg. Syst.}, 2015.

\bibitem{kleinberg1999authoritative}
Jon~M. Kleinberg and Jon M.
\newblock {Authoritative sources in a hyperlinked environment}.
\newblock {\em J. ACM}, 1999.

\bibitem{lin2013identifying}
Ying-Dar Lin, Yuan-Cheng Lai, Chien-Hung Chen, and Hao-Chuan Tsai.
\newblock {Identifying android malicious repackaged applications by
  thread-grained system call sequences}.
\newblock {\em Comput. Secur.}, 2013.

\bibitem{nadji2013connected}
Yacin Nadji, Manos Antonakakis, Roberto Perdisci, and Wenke Lee.
\newblock {Connected Colors: Unveiling the Structure of Criminal Networks}.
\newblock In {\em Res. Attacks, Intrusions, Defenses 16th Int. Symp.} Berlin,
  Heidelberg, 2013.

\bibitem{nidhi2012comparative}
Ritika~Wason Nidhi~Grover.
\newblock {Comparative Analysis Of Pagerank And HITS Algorithms}.
\newblock {\em Int. J. Eng. Res. Technol.}, 2012.

\bibitem{panjwani2015experimental}
Susmit Panjwani, Stephanie Tan, Keith~M. Jarrin, and Michel Cukier.
\newblock {An experimental evaluation to determine if port scans are precursors
  to an attack}.
\newblock In {\em Proc. Int. Conf. Dependable Syst. Networks}, 2005.

\bibitem{sanz2014anomaly}
Borja Sanz, Igor Santos, Xabier Ugarte-Pedrero, Carlos Laorden, Javier Nieves,
  and Pablo~Garcia Bringas.
\newblock {Anomaly detection using string analysis for android malware
  detection}.
\newblock In {\em Int. Jt. Conf. SOCO'13-CISIS'13-ICEUTE'13}, 2014.

\bibitem{spreitzenbarth2013mobile}
Michael Spreitzenbarth, Felix~C. Freiling, Florian Echtler, Thomas Schreck, and
  Johannes Hoffmann.
\newblock {Mobile-sandbox: having a deeper look into android applications}.
\newblock In {\em Proc. 28th Annu. ACM Symp. Appl. Comput.}, 2013.

\bibitem{suarez2014dendroid}
Guillermo Suarez-Tangil, Juan~E Tapiador, Pedro Peris-Lopez, and Jorge Blasco.
\newblock {Dendroid: A text mining approach to analyzing and classifying code
  structures in Android malware families}.
\newblock {\em Expert Syst. Appl.}, 2014.

\bibitem{sun2015droideagle}
Mingshen Sun, Mengmeng Li, and John C~S Lui.
\newblock {DroidEagle: seamless detection of visually similar Android apps}.
\newblock In {\em ACM Conf. Secur. Priv. Wirel. Mob. Networks}, WiSec '15,
  2015.

\bibitem{vidas2014a5}
Timothy Vidas, Jiaqi Tan, Jay Nahata, Chaur~Lih Tan, Nicolas Christin, and
  Patrick Tague.
\newblock {A5: Automated analysis of adversarial android applications}.
\newblock In {\em ACM Work. Secur. Priv. Smartphones Mob. Devices}, 2014.

\bibitem{wei2012android}
Te-En~En Wei, Ching-Hao~Hao Mao, Albert~B. Jeng, Hahn-Ming~Ming Lee,
  Horng-Tzer~Tzer Wang, and Dong-Jie~Jie Wu.
\newblock {Android malware detection via a latent network behavior analysis}.
\newblock In {\em Trust. Secur. Priv. Comput. Commun.}, 2012.

\bibitem{weimer2005passive}
Florian Weimer.
\newblock {Passive DNS Replication}.
\newblock Technical report, 2005.

\bibitem{yang2014apklancet}
Wenbo Yang, Juanru Li, Yuanyuan Zhang, Yong Li, Junliang Shu, and Dawu Gu.
\newblock {APKLancet: tumor payload diagnosis and purification for android
  applications}.
\newblock In {\em ACM Symp. Information, Comput. Commun. Secur.}, ASIA CCS '14,
  2014.

\bibitem{yuan2014droid}
Zhenlong Yuan, Yongqiang Lu, Zhaoguo Wang, and Yibo Xue.
\newblock {Droid-Sec: deep learning in android malware detection}.
\newblock In {\em ACM SIGCOMM Comput. Commun. Rev.}, 2014.

\bibitem{zhang2013vetting}
Yuan Zhang, Min Yang, Bingquan Xu, Zhemin Yang, Guofei Gu, Peng Ning, X~Sean
  Wang, and Binyu Zang.
\newblock {Vetting undesirable behaviors in android apps with permission use
  analysis}.
\newblock In {\em ACM SIGSAC Conf. Comput. Commun. Secur.}, 2013.

\bibitem{zhongyang2013droidalarm}
Yibing Zhongyang, Zhi Xin, Bing Mao, and Li~Xie.
\newblock {DroidAlarm: An All-sided Static Analysis Tool for Android
  Privilege-escalation Malware}.
\newblock In {\em ACM SIGSAC Symp. Information, Comput. Commun. Secur.}, ASIA
  CCS '13, 2013.

\bibitem{zhou2013fast}
Wu~Zhou, Yajin Zhou, Michael Grace, Xuxian Jiang, and Shihong Zou.
\newblock {Fast, scalable detection of piggybacked mobile applications}.
\newblock In {\em ACM Conf. Data Appl. Secur. Priv.}, CODASPY '13, 2013.

\bibitem{zhou2012detecting}
Wu~Zhou, Yajin Zhou, Xuxian Jiang, and Peng Ning.
\newblock {Detecting Repackaged Smartphone Applications in Third-party Android
  Marketplaces}.
\newblock In {\em Proc. Second ACM Conf. Data Appl. Secur. Priv.}, 2012.

\bibitem{zhou2012hey}
Yajin Zhou, Zhi Wang, Wu~Zhou, and Xuxian Jiang.
\newblock {Hey, You, Get Off of My Market: Detecting Malicious Apps in Official
  and Alternative Android Markets.}
\newblock In {\em Symp. Netw. Distrib. Syst. Secur.}, 2012.

\end{thebibliography}

\end{document}